\documentstyle[12pt,epsfig]{article}

\catcode`\@=11
\def\@citex[#1]#2{%
\if@filesw \immediate \write \@auxout {\string \citation {#2}}\fi
\@tempcntb\m@ne \let\@h@ld\relax \def\@citea{}%
\@cite{%
  \@for \@citeb:=#2\do {%
    \@ifundefined {b@\@citeb}%
      {\@h@ld\@citea\@tempcntb\m@ne{\bf ?}%
      \@warning {Citation `\@citeb ' on page \thepage \space undefined}}%
      {\@tempcnta\@tempcntb \advance\@tempcnta\@ne%
      \@tempcntb\number\csname b@\@citeb \endcsname \relax%
      \ifnum\@tempcnta=\@tempcntb 
        \ifx\@h@ld\relax%
          \edef \@h@ld{\@citea\csname b@\@citeb\endcsname}%
        \else%
          \edef\@h@ld{\ifmmode{-}\else--\fi\csname b@\@citeb\endcsname}%
        \fi%
      \else
        \@h@ld\@citea\csname b@\@citeb \endcsname%
        \let\@h@ld\relax%
      \fi}%
    \def\@citea{,\penalty\@highpenalty\,}%
  }\@h@ld
}{#1}}

\def\@citeb#1#2{{[#1]\if@tempswa , #2\fi}}
\def\@citeu#1#2{{$^{#1}$\if@tempswa , #2\fi }}
\def\@citep#1#2{{#1\if@tempswa , #2\fi}}

\def\bcites{         
        \catcode`\@=11
        \let\@cite=\@citeb
        \catcode`\@=12
}

\def\upcites{         
        \catcode`\@=11
        \let\@cite=\@citeu
        \catcode`\@=12
}

\def\plaincites{      
        \catcode`\@=11
        \let\@cite=\@citep
        \catcode`\@=12
}

\newcount\hour
\newcount\minute
\newtoks\amorpm
\hour=\time\divide\hour by 60
\minute=\time{\multiply\hour by 60 \global\advance\minute by-\hour}
\edef\standardtime{{\ifnum\hour<12 \global\amorpm={am}%
        \else\global\amorpm={pm}\advance\hour by-12 \fi
        \ifnum\hour=0 \hour=12 \fi
        \number\hour:\ifnum\minute<10 0\fi\number\minute\the\amorpm}}
\edef\militarytime{\number\hour:\ifnum\minute<10 0\fi\number\minute}

\def\draftlabel#1{{\@bsphack\if@filesw {\let\thepage\relax
   \xdef\@gtempa{\write\@auxout{\string
      \newlabel{#1}{{\@currentlabel}{\thepage}}}}}\@gtempa
   \if@nobreak \ifvmode\nobreak\fi\fi\fi\@esphack}
        \gdef\@eqnlabel{#1}}
\def\@eqnlabel{}
\def\@vacuum{}
\def\marginnote#1{}
\def\draftmarginnote#1{\marginpar{\raggedright\scriptsize\tt#1}}
\def\draft{
        \pagestyle{plain}
        \overfullrule=2pt
        \oddsidemargin -.5truein
        \def\@oddhead{\sl \phantom{\today\quad\militarytime} \hfil
        \smash{\Large\sl DRAFT} \hfil \today\quad\militarytime}
        \let\@evenhead\@oddhead
        \let\label=\draftlabel
        \let\marginnote=\draftmarginnote
        \def\ps@empty{\let\@mkboth\@gobbletwo
        \def\@oddfoot{\hfil \smash{\Large\sl DRAFT} \hfil}
        \let\@evenfoot\@oddhead}
        \def\@eqnnum{(\theequation)\rlap{\kern\marginparsep\tt\@eqnlabel}%
        \global\let\@eqnlabel\@vacuum}  }

\def\blackfonts{
        \font\blackboard=msbm10 scaled\magstep1
        \font\blackboards=msbm8
        \font\blackboardss=msbm6
}

\def\nblack{            
        \def\ZZ{{Z \n{10} Z}}
        \def\NN{{N \n{14} N}}
        \def\CC{{C \n{11} C}}
        \def\RR{{R \n{11} R}}
        \def\QQ{{Q \n{12} Q}}
        \def\PP{{P \n{11} P}}
}

\def\prep{         
        \catcode`\@=11
        \input art10.sty
        \catcode`\@=12
        
        \let\small\null
        \def\blackfonts{
                \font\blackboard=msbm10
                \font\blackboards=msbm7
                \font\blackboardss=msbm5
        }
        \let\sl\it
        \twocolumn
        \sloppy
        \voffset=-2.54truecm
        \hoffset=-2.54truecm
        \flushbottom
        \parindent 1em
        \leftmargini 2em
        \leftmarginv .5em
        \leftmarginvi .5em
        \marginparwidth 48pt
        \marginparsep 10pt
        \setlength{\columnsep}{2truecm}
        \setlength{\textwidth}{25.4truecm}
        \setlength{\textheight}{17truecm}
        \baselineskip=16pt
        \oddsidemargin .18truein
        \evensidemargin .17truein
}

\def\eqalign#1{\null\,\vcenter{\openup\jot\m@th
  \ialign{\strut\hfil$\displaystyle{##}$&$\displaystyle{{}##}$\hfil
      \crcr#1\crcr}}\,}
\def\eqalignno#1{\displ@y \tabskip\centering
  \halign to\displaywidth{\hfil$\@lign\displaystyle{##}$\tabskip\z@skip
    &$\@lign\displaystyle{{}##}$\hfil\tabskip\centering
    &\llap{$\@lign##$}\tabskip\z@skip\crcr
    #1\crcr}}

\def\section{\@startsection {section}{1}{\z@}{3.ex plus 1ex minus
 .2ex}{2.ex plus .2ex}{\large\bf}}
\def\subsection{\@startsection{subsection}{2}{\z@}{2.75ex plus 1ex minus
 .2ex}{1.5ex plus .2ex}{\bf}}        

\def\appendix{{\newpage\section*{Appendix}}\let\appendix\section%
        {\setcounter{section}{0}
        \gdef\thesection{\Alph{section}}}\section}

\def\abstract{\if@twocolumn
\section*{Abstract}
\else 
\begin{center}
{\bf Abstract\vspace{-.5em}\vspace{0pt}}
\end{center}
\quotation
\fi}

\catcode`\@=12

\newcommand{\beq}{\begin{equation}}
\newcommand{\eeq}{\end{equation}}
\newcommand{\beqa}{\begin{eqnarray}}
\newcommand{\eeqa}{\end{eqnarray}}

\newcommand{\Z}{{\bf Z}}
\newcommand{\Q}{{\bf Q}}
\newcommand{\R}{{\bf R}}
\newcommand{\C}{{\bf C}}
\newcommand{\e}{{\rm e}}

\newcommand{\dd}{{\rm d}}

\def\noj#1,#2,{{\bf #1} (19#2)\ }
\def\jou#1,#2,#3,{{\sl #1\/ }{\bf #2} (19#3)\ }
\def\ann#1,#2,{{\sl Ann.\ Physics\/ }{\bf #1} (19#2)\ }
\def\cmp#1,#2,{{\sl Comm.\ Math.\ Phys.\/ }{\bf #1} (19#2)\ }
\def\ma#1,#2,{{\sl Math.\ Ann.\/ }{\bf #1} (19#2)\ }
\def\ng#1,#2,{{\sl Nagoya.\ Math.\ J.\/ }{\bf #1} (19#2)\ }
\def\jd#1,#2,{{\sl J.\ Diff.\ Geom.\/ }{\bf #1} (19#2)\ }
\def\invm#1,#2,{{\sl Invent.\ Math.\/ }{\bf #1} (19#2)\ }
\def\cq#1,#2,{{\sl Class.\ Quantum Grav.\/ }{\bf #1} (19#2)\ }
\def\cqg#1,#2,{{\sl Class.\ Quantum Grav.\/ }{\bf #1} (19#2)\ }
\def\ijmp#1,#2,{{\sl Int.\ J.\ Mod.\ Phys.\/ }{\bf A#1} (19#2)\ }
\def\jmphy#1,#2,{{\sl J.\ Geom.\ Phys.\/ }{\bf #1} (19#2)\ }
\def\jams#1,#2,{{\sl J.\ Amer.\ Math.\ Soc.\/ }{\bf #1} (19#2)\ }
\def\grg#1,#2,{{\sl Gen.\ Rel.\ Grav.\/ }{\bf #1} (19#2)\ }
\def\mpl#1,#2,{{\sl Mod.\ Phys.\ Lett.\/ }{\bf A#1} (19#2)\ }
\def\nc#1,#2,{{\sl Nuovo Cim.\/ }{\bf #1} (19#2)\ }
\def\np#1,#2,{{\sl Nucl.\ Phys.\/ }{\bf B#1} (19#2)\ }
\def\pl#1,#2,{{\sl Phys.\ Lett.\/ }{\bf #1B} (19#2)\ }
\def\pla#1,#2,{{\sl Phys.\ Lett.\/ }{\bf #1A} (19#2)\ }
\def\pr#1,#2,{{\sl Phys.\ Rev.\/ }{\bf #1} (19#2)\ }
\def\prd#1,#2,{{\sl Phys.\ Rev.\/ }{\bf D#1} (19#2)\ }
\def\prl#1,#2,{{\sl Phys.\ Rev.\ Lett.\/ }{\bf #1} (19#2)\ }
\def\prp#1,#2,{{\sl Phys.\ Rept.\/ }{\bf #1C} (19#2)\ }
\def\ptp#1,#2,{{\sl Prog.\ Theor.\ Phys.\/ }{\bf #1} (19#2)\ }
\def\ptpsup#1,#2,{{\sl Prog.\ Theor.\ Phys.\/ Suppl.\/ }{\bf #1} (19#2)\ }
\def\rmp#1,#2,{{\sl Rev.\ Mod.\ Phys.\/ }{\bf #1} (19#2)\ }
\def\yadfiz#1,#2,#3[#4,#5]{{\sl Yad.\ Fiz.\/ }{\bf #1} (19#2) #3%
\ [{\sl Sov.\ J.\ Nucl.\ Phys.\/ }{\bf #4} (19#2) #5]}
\def\zh#1,#2,#3[#4,#5]{{\sl Zh.\ Exp.\ Theor.\ Fiz.\/ }{\bf #1} (19#2) #3%
\ [{\sl Sov.\ Phys.\ JETP\/ }{\bf #4} (19#2) #5]}

\def\beq{\begin{equation}}
\def\eeq{\end{equation}}
\def\beqar{\begin{eqnarray}}
\def\eeqar{\end{eqnarray}}

\newcommand{\be}{\begin{equation}}
\newcommand{\ee}{\end{equation}}
\newcommand{\bea}{\begin{eqnarray}}
\newcommand{\eea}{\end{eqnarray}}

\def\nfrac#1#2{{\displaystyle{\vphantom1\smash{\lower.5ex\hbox{\small$#1$}}%
        \over\vphantom1\smash{\raise.25ex\hbox{\small$#2$}}}}}

\def\n#1{\mskip-#1mu}

\def\to{\rightarrow}

\def\lae{\mathrel{\mathop{\smash{\lower .5 ex \hbox{$\stackrel<\sim$}}}}}
\def\lae{\mathrel{\mathop{\smash{\lower .5 ex \hbox{$\stackrel>\sim$}}}}}

\def\l:{\mathopen{:}\,}
\def\r:{\,\mathclose{:}}

\def\la{\langle}
\def\ra{\rangle}

\catcode`\@=11
\def\theequation{\arabic{equation}}
\catcode`\@=12

\nblack
\bcites

\nblack

\catcode`\@=11
\def\theequation{\thesection.\arabic{equation}}
\@addtoreset{equation}{section}
\@addtoreset{footnote}{section}
\@addtoreset{footnote}{subsection}
\catcode`\@=12

\newcommand{\beqn}{\begin{equation}}
\newcommand{\eeqn}{\end{equation}}
\newcommand{\beqnarray}{\begin{eqnarray}}
\newcommand{\eeqnarray}{\end{eqnarray}}
\newcommand {\bear} [1] {\begin {array} {#1}}
\newcommand {\ear} {\end {array}}

\newcommand {\beqarn} {\begin{eqnarray*}}
\newcommand {\eeqarn} {\end{eqnarray*}}

\begin{document}

\begin{titlepage}

\begin{center}
\today
\hfill LBNL-42658, UCB-PTH-99/02\\
\hfill                  hep-th/9902102

\vskip 1.5 cm
{\large \bf D-Branes, T-Duality, and Index Theory}
\vskip 1 cm 
{Kentaro Hori}\\
\vskip 0.5cm
{\sl Department of Physics,
University of California at Berkeley\\
366 Le\thinspace Conte Hall, Berkeley, CA 94720-7300, U.S.A.\\
and\\
Theoretical Physics Group, Mail Stop 50A--5101\\
Ernest Orlando Lawrence Berkeley National Laboratory\\
Berkeley, CA 94720, U.S.A.\\}

\end{center}

\vskip 0.5 cm
\begin{abstract}
We show that the transformation of D-branes under T-duality on
four-torus is represented by Nahm transform of instantons. The
argument for this allows us to generalize Nahm transform to the case
of orthogonal and symplectic gauge groups as well as to instantons
on $\Z_2$ orbifold of four-torus. In addition, we identify the
isomorphism of K-theory groups that realizes the transformation of
D-brane charges under T-duality on torus of arbitrary dimensions.
By the isomorphism we are lead to identify the correct K-theory group
that classifies D-brane charges in Type II orientifold.

\end{abstract}

\end{titlepage}

\newcommand{\hT}{\widehat{T}}
\newcommand{\Dsl}{D\!\!\!\! /\,}
\newcommand{\hx}{\hat{x}}
\newcommand{\Ker}{{\rm Ker}}
\newcommand{\K}{{\rm K}}
\newcommand{\tilK}{\widetilde{\rm K}}
\newcommand{\KO}{{\rm KO}}
\newcommand{\KSp}{{\rm KSp}}
\newcommand{\KR}{{\rm KR}}
\newcommand{\KPR}{{\rm KpR}}
\newcommand{\KZt}{{\rm K}_{\Z_2}}
\newcommand{\ind}{{\rm ind}\,}
\newcommand{\ch}{{\rm ch}}
\newcommand{\D}{{\cal D}}
\newcommand{\rank}{{\rm rank}}

\section{Introduction}

A D$p$-brane wrapped
on a circle is mapped under T-duality to a D$(p-1)$-brane
where the $U(1)$ Wilson line of the D$p$-brane
corresponds to the position of the D$(p-1)$-brane
in the dual circle.
This was known since the discovery of D-branes \cite{DLP,Horava1}
and this is actually how D-branes are discovered.
However, the T-duality transformation rule is less obvious
for more general configuration of D-branes in a more general
string background.

In superstring theory, D-branes are sources of Ramond-Ramond (RR)
potentials \cite{Polchinski}.
Transformation of RR fields under T-duality was studied
in \cite{BHO,EJL}.
RR fields of the theory on $T^n\times M$
and those of the T-dual theory on $\hT^n\times M$
are related as
\beq
\widehat{\!\!\vbox to 11pt{}\e^{-B}F\,\,}
=\int\limits_{{\vbox to 6pt{}T}^n}{\rm ch}({\cal P})\,\e^{-B}F.
\label{RRflux}
\eeq
Here, $B$ is the Neveu-Schwarz B-field and
 $F=\sum_pF_{p+2}$ is the sum of gauge invariant
RR field strengths where the sum is over $p=0,2,4,\ldots$ for Type IIA
and $p=-1,1,3,\ldots$ for Type IIB.\footnote{
We choose the following field definitions.
The gauge transformation for the R-R potential $A=\sum A_{p+1}$
is given by $\delta A=\e^B\dd \alpha$ for $\alpha=\sum\alpha_p$,
and therefore the gauge invariant field strength is
$F=\e^{B}\dd(\e^{-B}A)=\dd A-H\wedge A$ subject to the Bianchi identity
$\dd F=H\wedge F$.}
Also, ${\rm ch}({\cal P})=\exp(\sum_{i=1}^n\dd \widehat{t}_i\wedge
\dd t^i)$ in which $t^i$ and $\widehat{t}_i$ are coordinates
of $T^n$ and $\hT^n$ dual to each other.
As the name suggests, ${\rm ch}({\cal P})$
represents the Chern character of
a complex line bundle over $\hT^n\times T^n$
--- the so called Poincar\'e bundle ${\cal P}$
--- which has a $U(1)$ connection with curvature
$-2\pi i\sum_i\dd \widehat{t}_i\wedge \dd t^i$.

The simplicity of the transformation rule
(\ref{RRflux}) is intriguing and begs for an explanation
which is independent of the supergravity or worldsheet computation.
It is also interesting to ask whether similar rule exists 
for more general cases, such as the case with orbifold or orientifold
projection.
We note here that (\ref{RRflux}) looks like a formula
which appears in the family index theory \cite{ASIV,ASV,ASAd};
in particular,
for a family of Dirac operators on $T^n$ parametrized by
$\hT^n$
which is carried by a bundle over $T^n\times \hT^n$
determined by ${\cal P}$.
Since D-branes are RR sources,
this suggests that there is
a similar index-theoretic transformation rule
for D-branes, which might lead to a simplest explanation
for (\ref{RRflux}) and to its generalization.
The purpose of this paper is to
find such a transformation rule for D-brane configurations.

Actually, it has been suspected that an index-theoretic
transformation realizes T-duality on four-torus $T^4$
(and also $T^n$ with $n=1,2,3$) in the case
where there are D$p$-branes at points on $T^4$
within wrapped D$(p+4)$-branes.
In such a case, a D-brane configuration can be represented as
an instanton configuration on $T^4$, and it was
suggested in \cite{DoMo}
that T-duality may possibly be
realized as
a transform of
representing instanton configurations
called {\it Nahm transform} \cite{Sh,BvB,DK}
(or {\it Mukai's Fourier transform} \cite{Mu}).
Nahm transform uses a family of Dirac operators on $T^4$ parametrized
by $\hT^4$ which is carried by the Poincar\'e bundle
${\cal P}$, and
the transformation of the topological numbers of the instantons
is given by an index formula which looks like (\ref{RRflux}).
In particular,
the rank and the instanton number are interchanged.
In terms of D-brane charges, this corresponds to
the interchange of $(p+4)$-brane and $p$-brane charges
which is exactly what T-duality does.
However,
there was no other argument showing why the Nahm transform
can be identified with the T-duality on four-torus.
In this paper, we provide a simple argument which shows
that one can indeed identify the Nahm transform and T-duality.
The argument is so simple that it can be easily generalized to
the case with orbifold/orientifold projection, finding
a generalization of the Nahm transform to, say,
orthogonal/symplectic gauge groups.

Based on recent analysis of brane-anti-brane systems (for example in
\cite{Sen}),
it was argued by Witten \cite{WittenK} that,
as has been proposed in \cite{MinaMoo},
D-brane charge takes values in an appropriate
K-theory group of the space-time $X$ when $X$ is large
compared to the string scale.
(See \cite{Horava,Garcia,Gukov} for subsequent studies.)
If this also holds for any size of $X$,
T-duality should induce an isomorphism of a K-theory group of
$T^n\times M$ to another K-theory group of
the dual $\hT^n\times M$.
In fact, brane-anti-brane systems naturally appear in
our argument for ``Nahm transform = T-duality'',
and an isomorphism between K-theory groups emerges very naturally.
This seems to be actually not limited to four-torus;
As another non-trivial application of our argument,
we will construct the isomorphism of K-theory groups
for T-duality on torus of other dimensions.
During the course, we find the identification
of the K-theory group for Type II $\Z_2$-orientifold:
D-brane charges in the presence of
an orientifold $p$-plane are classified by
$\KR^{-(9-p)}$ or $\KR^{-(5-p)}$
for $SO$- or $Sp$-type orientifold respectively.

This paper is organized as follows.
In section 2, we give a simple argument which shows that
Nahm-transform can indeed
be identified as the transform of D-branes under T-duality on
four-torus. We extend the argument to the case where the Nahm
transform fails to become a vector bundle and give the interpretation
of the resulting object as a certain bound state of D-branes and
anti-D-branes.
The basic argument of section 2 is applied in section 3
to string theory with orientifold or orbifold projections.
This leads to the generalization of Nahm transform to
instantons of orthogonal and symplectic gauge groups as well as to
instantons on $\Z_2$ orbifold.
In section 4, as an application of
the picture of brane-anti-brane bound state,
we obtain index-theoretic isomorphisms of K-theory groups
that realizes T-duality on four-torus.
This is extended in section 5 to find isomorphisms of K-theory groups
for T-duality on torus of other dimensions.
There we also indentify the K-theory group
that classifies D-branes in Type II orientifold.

T-duality was discussed in terms of
K-theory also in
\cite{Sharpe,BGH}
which appeared after this paper.
The former considers D-branes in algebraic varieties
while the latter takes a close look at
T-duality in Type II orientifold.
They are closely related to sections 2 and 5 respectively.

\section{D-branes on Four-Torus and T-duality}

In this section we study how D-brane configurations
are transformed under T-duality on four-torus in Type II string
theory. 

Let us consider Type II string theory on $\R^6\times T^4$
with $N$ D$(p+4)$-branes wrapped on $T^4$.
We choose flat orthonormal space-time coordinates $x^M=x^0,\ldots,x^9$
and consider $T^4$ as $\R^4/\Lambda$ where $\R^4$
is spanned by
$x^{\mu}=x^1,x^2,x^3,x^4$ and $\Lambda=\{n^{\mu}\}\subset \R^4$ is
a lattice of rank four.
The theory contains $U(N)$ super-Yang-Mills theory on the brane
with sixteen supercharges.
The supersymmetry transformation of the gaugino $\psi$ is given by
\beq
\delta \psi=-{1\over 4}F_{MN}\Gamma^{MN}\epsilon
\label{SUSY}
\eeq
where $\epsilon$ is an anticommuting parameter which is a
positive chirality spinor in ten dimensions
and $F_{MN}$ is the field strength of
the (extended) $U(N)$ gauge field.
If the gauge field
has a self-dual field strength
on $T^4$ with non-zero instanton number, 
the supersymmetry is broken to half and only those with
positive chirality on $T^4$
are unbroken.
This is because $F_{\mu\nu}\Gamma^{\mu\nu}={1\over 2}
F^{\mp}_{\mu\nu}\Gamma^{\mu\nu}$ on spinors with chirality $\pm$
on $T^4$ where $F^{\pm}=F\pm *F$.
\footnote{In this paper, positive or $+$ (resp. negative or $-$)
chirality in four dimensions means $\Gamma^{1234}=1$ (resp. $-1$)
(and likewise, in 1+1 or 5+1 dimensions (spanning 09 or 012349
directions), it means $\Gamma^{09}$ or $\Gamma^{012349}$ $=1$
(resp. $-1$)). This convention in four-dimensions appears to
conflict with mathematical notation where in
$2n$-dimension spinor bundles with $i^n\Gamma^{12\cdots 2n}=
\pm 1$ is denoted as $S^{\pm}$. Nevertheless, we shall follow this
mathematical convention as far as the notation $S^{\pm}$ is concerned.
Thus, in this paper the bundle of {\it positive} (or +) chirality
spinors (with $\Gamma^{1234}=1$) is denoted as $S^-$ whereas $S^+$
is the bundle of {\it negative} (or $-$) chirality spinors
(with $\Gamma^{1234}=-1$). I hope that this does not confuse the
reader.}
This together with the fact that an instanton on D$(p+4)$-branes
carries a unit charge of the RR $p$-form potential through the
Chern-Simons coupling \cite{BinB} suggest that the instantons on
$T^4$ can be identified with D$p$-branes at points of $T^4$.
Namely, $N$ D$(p+4)$-branes with a $k$-instanton configuration on
$T^4$ can be considered as a classical bound state of
$N$ D$(p+4)$-branes and $k$ D$p$-branes.\footnote{It was noted
in \cite{HarveyMoore} that
(\ref{SUSY}) admits another term
with values in the center of $U(N)$
which relaxes the
self-duality condition for supersymmetry.
In this section we do not consider
the associated generalization which necessarily involves
the first Chern class or D$(p+2)$-branes.}
This description of D$p$-branes as instantons on D$(p+4)$-branes
 is of course valid only when the size of the torus is large
compared to the string scale.

Now we address the following question:
How is T-duality transformation of D-branes described in terms of
gauge fields on $T^4$? We first have to note that T-duality
inverts the size of the torus (when B-field is zero) with respect to
the string scale. Thus, the question makes sense only when
there is a natural one-to-one correspondence between
D-brane configurations on large and small torus.
In the present case, eight supersymmetries and
non-renormalization theorm assures this,
since parameters for D-brane configuration and
the size of the torus belong to different supermultiplets.

In what follows we provide an answer to this question.
We will work in the case $p=5$ since we want to introduce
branes of lower dimensions as probes.
This means that we work in
(unphysical) Type IIB string theory with
$N$ D9-branes on $\R^6\times T^4$.
Tadpole, anomaly and other sickness of the theory
do not affect our argument which we are going to make.
We could also work in the barely physical case of $p=4$
in which most of our argument can be repeated,
although it is less convenient compared to $p=5$.

\subsection{Nahm Transform as T-duality}

\subsection*{\sl Probing by a D1-brane}

Let us first probe this D9-brane system by a D1-brane which
spans the coordinates $x^{0,9}$, or $x^{\pm}=x^0\pm x^9$.
The analysis of lowlying spectrum of open strings ending
on this probe (in the case where $T^4$ is large)
is the same as the one for a D1-brane in
Type I string theory \cite{PW}
except that we do not impose invariance under worldsheet orientation
reversal.
The theory on the probe is a $U(1)$ gauge theory in $1+1$ dimensions with
at most $(0,8)$ supersymmetry. It has an $(8,8)$ $U(1)$ vector
multiplet whose scalar components take values in $\R^4\times T^4$
and also a positive-chirality fermion $\lambda$ with
$U(1)$ charge $-1$
which transforms in the fundamental representation of the flavor group
$U(N)$. The fermion $\lambda$ is coupled to the $U(N)$ gauge field
$A_{\mu}\dd x^{\mu}$
on the D9-branes via the minimal coupling
\beq
\bar\lambda\,
\Bigl(\,\partial_- +\partial_-X^{\mu}A_{\mu}(X)-ia_-\,\Bigr)\,
\lambda
\eeq
where $ia_{\pm}\dd x^{\pm}$
is the $U(1)$ gauge field and $X^{\mu}$ are the scalar
fields in the $U(1)$ vector multiplet representing the position of
the probe in $T^4$.
If the gauge field $A_{\mu}$ is flat, the theory preserves the
$(0,8)$ supersymmetry
(generated by those $\epsilon$ with $\Gamma^0\Gamma^1=1$)
but if it is in an instanton configuration with
self-dual curvature, $F^-_{\mu\nu}=0$, the supersymmetry is broken to
$(0,4)$ (generated by those $\epsilon$ with $\Gamma^0\Gamma^9=
\Gamma^1\Gamma^2\Gamma^3\Gamma^4=1$).
For a finite size torus, there are actually infinitely many
additional modes coming from the strings winding around 1-cycles
of $T^4$, but the effect of them is small when the size of $T^4$
is large and they simply decouple in the low energy limit.

\subsection*{\sl Probing by a Wrapped D5-brane}

Next, let us probe the D9-brane system by a D5-brane wrapped on $T^4$
and spanning the $x^{0,9}$ or $x^{\pm}$ directions.
The analysis of low-lying modes on the D5-brane
is almost identical to that in \cite{small}.
The theory of these modes is
a $U(1)$ gauge theory in $5+1$ dimensions with at most
$(1,0)$ supersymmetry. It has a $(1,1)$ $U(1)$ vector multiplet
and a $(1,0)$ hypermultiplet with $U(1)$ charge $-1$ which
transforms in the fundamental representation of the flavor group
$U(N)$.
The flavor group $U(N)$ has a
background gauge field which is given by
the gauge field $A_{\mu}\dd x^{\mu}$ of the D9-brane.
Namely, the hypermultiplet fields
--- scalars $Q^{\sigma}$ ($\sigma=1,2$) and a negative chirality
spinor $\Psi$ --- are minimally coupled to $A_{\mu}$
via
\beq
\parallel \!D Q^{\sigma}\!\parallel^2
+
\,\overline{\Psi} \,D\!\!\!\! / \,\,\Psi\,,
\label{hypkin}
\eeq
where
$D_{\pm}=\partial_{\pm}-ia_{\pm}$
and
\beq
D_{\mu}=\partial_{\mu}+A_{\mu}-ia_{\mu}\,,
\label{defDmu}
\eeq
in which $ia_{\pm}\dd x^{\pm}+ia_{\mu}\dd x^{\mu}$
is the $U(1)$ gauge field of the probe.
When the $U(N)$ gauge field $A_{\mu}\dd x^{\mu}$ is flat,
the theory preserves the
$(1,0)$ supersymmetry
but if it is in an instanton configuration on $T^4$ with
self-dual curvature $F^-_{\mu\nu}=0$,
the supersymmetry is broken to half
--- only those with positive-chirality in $T^4$ (as well as in
$\R^2\times T^4$) is unbroken.
We shall consider the case where
the instanton number is $k$, namely $A_{\mu}$ is a gauge
field that defines a connection of a rank $N$ complex vector bundle $E$
over $T^4$ with the second Chern character ${\rm ch}_2(E)=k$.
Then,
the wrapped D5-brane is probing a bound state of
$N$ D9-branes wrapped on $T^4$ and $k$ D5-branes at points in $T^4$.

When the size of $T^4$ becomes small, it is natural to consider the
theory of the wrapped D5-brane probe as $1+1$ dimensional theory
with a tower of infinite Kaluza-Klein modes.
The theory is invariant under $(0,4)$ supersymmetry which is
generated by those $\epsilon$ with
$\Gamma^0\Gamma^9=\Gamma^1\Gamma^2\Gamma^3\Gamma^4=1$.
At long distances, massive modes simply decouple and
only finitely many massless modes remain in the theory. 
From the six-dimensional $(1,1)$ $U(1)$ vector multiplet, we obtain 
$(0,4)$ multiplets whose bosonic fields are
the $U(1)$ gauge field $a_{\pm}$ together with
the scalar fields taking
values in the space of solutions to the equations
\beqa
&&\partial_{\mu}X^p=0,\\
&&\partial_{\mu}a_{\nu}-\partial_{\nu}a_{\mu}=0,\label{flat}
\eeqa
where $X^p$ ($p=5,6,7,8$) are the scalar component of the $(1,1)$
vector multiplet.
The solution space for the first equation is $\R^4$ since it
is solved by $X^p=$ constant.
The second equation is the equation for $ia_{\mu}\dd x^{\mu}$ to be
a flat $U(1)$ connection over $T^4$.
Since we have a $U(1)$ gauge symmetry, what we actually obtain as
the target space is the {\it moduli space} of flat $U(1)$ connections.
A flat $U(1)$ gauge field can have
$a_{\mu}=$ constant, and $a_{\mu}$ is gauge equivalent to
$a_{\mu}+\hat{n}_{\mu}$ where $\hat{n}_{\mu}$ belongs to the dual
lattice $2\pi\Lambda^*$ of $\Lambda/2\pi$ so that the gauge
transformation $g(x)=\e^{i\hat{n}x}$ is single valued on $T^4$.
The moduli space is therefore the dual torus
$\hT^4=\{a_{\mu}\}/\{\hat{n}_{\mu}\}=(\R^4)^*/2\pi\Lambda^*$.
Thus, from the six-dimensional (1,1) $U(1)$ vector multiplet,
we obtain a $1+1$ dimensional $(8,8)$ $U(1)$ vector multiplet
whose scalar components take values in $\R^4\times \hT^4$.

Massless fields in $1+1$ dimensions
also come from the hypermultiplet in the bifundamental representation
and these break the
supersymmetry to $(0,4)$.
They correspond to the fields on $T^4$ satisfying
\beq
D_{\mu}Q^{\sigma}=0,~~~
\gamma^{\mu}D_{\mu}\psi=0\,,
\label{zerom}
\eeq
where $D_{\mu}$ is given by (\ref{defDmu}).
For a generic instanton configuration, there is no covariantly
constant section of the associated bundle in the
fundamental representation.
Therefore, there is no solution to $D_{\mu}Q^{\sigma}=0$.
In this case, there is neither positive-chirality Dirac zero mode
since such a thing, if existed, would be covariantly constant 
in a self-dual instanton background
as one can see from the Weizenbock formula
$\Dsl^{\dag}\Dsl=D^{\dag}D -F_{\mu\nu}\gamma^{\mu\nu}$.
However, as the index theorem shows,
there are $k$ Dirac zero modes with negative
chirality (see section 4).
Let us choose a family of orthonormal basis
$\psi_1(a),\ldots,\psi_k(a)$ of the space of zero modes
which varies smoothly as $a_{\mu}$ is varied.
Let us expand the field $\Psi$ as
\beq
\Psi(x^{\pm},x^{\mu})
=\sum_{i=1}^k\psi_i(a(x^{\pm}))(x^{\mu})\otimes\lambda^i(x^{\pm}),
\eeq
where
$\lambda_i$ are positive chirality spinors in $1+1$ dimensions
(positive
because $\Psi$ and $\psi_i(a)$ are negative
in $5+1$ and $4$-dimensions respectively).
Inserting this in the lagrangian (\ref{hypkin})
and integrating over $T^4$, we obtain a lagrangian for
$\lambda=(\lambda^i)$
\beq
\bar\lambda\,
\Bigl(\,\partial_-+\partial_-a_{\mu}\hat{A}^{\mu}(a)-ia_-\,\Bigr)\,
\lambda,
\eeq
where
\beq
\hat{A}^{\mu}(a)_{\bar\imath j}
=\int\limits_{T^4}\dd^4 x\,
\psi_i(a)^{\dag}{\partial\over\partial a_\mu}
\psi_j(a).
\label{NA}
\eeq
This $\hat{A}^{\mu}(a)$ can naturally be considered as a $U(k)$
gauge field on the parameter space $(\R^4)^*=\{a_{\mu}\}$
since the choice of orthonormal basis
$\psi_i(a)$ is arbitrary and can be changed to other ones.
Thus, we have obtained from the hypermultiplet
a positive chirality fermion $\lambda$
with a unit $U(1)$ charge which is coupled to the external gauge field
$\hat{A}^{\mu}(a)$ given by (\ref{NA}).
In other words, $\lambda$ takes values in
the bundle $\hat{E}$ of Dirac zero modes
provided with the unitary connection $\hat{A}$;
the fibre of $\hat{E}$ at $a\in (\R^4)^*$ is the kernel of
$\Dsl$ associated with $D_{\mu}=\partial_{\mu}+A_{\mu}-ia_{\mu}$
which is provided with the natural inner product coming from
the integration over $T^4$.
Note that the $U(1)$ gauge transformation $\Psi\to\e^{i\hat{n}x}\Psi$
sends a zero mode $\psi$ at $a$ to a zero mode $\tilde{\psi}$
at $a+\hat{n}$
defined by $\tilde{\psi}(x)=\e^{i\hat{n}x}\psi(x)$.
Thus, $\hat{E}$ should be considered as a bundle over
$(\R^4)^*/2\pi\Lambda^*=\hT^4$
where we identify $\psi\equiv \tilde{\psi}$.
It is clear that $\hat{A}$ in (\ref{NA}) defines a connection on it.

We see that the 1+1 dimensional theory we obtained
is exactly the same as the effective
theory of a D1-brane in Type IIB string theory on $\R^6\times\hT^4$
which probes D9-branes supporting the bundle $\hat{E}$
with connection $\hat{A}$.

The transform $A\mapsto \hat{A}$ of a gauge field on
$T^4$ to that of its dual $\hT^4$ is nothing but what is
known as the Nahm transform \cite{Sh,BvB,DK}.
In particular,
the $U(k)$ gauge field $\hat{A}^{\mu}\dd a_{\mu}$ on $\hT^4$
is actually an instanton
with self-dual field strength $\widehat{F}=*\widehat{F}$,
as expected from supersymmetry.
As we will see in section 4 using the index theorem,
the topological numbers of the bundle $\hat{E}$ is related to that of
$E$ as
${\rm rank}(\hat{E})=\ch_2(E)=k$ and
$\ch_2(\hat{E})={\rm rank}(E)=N$.
Thus, the system of D9-branes wrapped on $\hT^4$
supporting $(\hat{E},\hat{A})$ can be considered as 
a system of $k$ D9-branes wrapped on $\hT^4$
and $N$ D5-branes at points on $\hT^4$.

To summarize, we have seen that the theory on the wrapped
D5-brane probe
is effectively the same as
the effective theory on a D1-brane moving in $\R^6\times\hT^4$
and probing the system of D9-branes
which support the gauge field configuration
$\hat{A}^{\mu}(a)\dd a_{\mu}$ on the dual torus $\hT^4$
given by (\ref{NA})
(which represent a system of $k$ D9
and $N$ D5 branes).
By definition, the effective theory of a D5-brane wrapped on a small
$T^4$ is identified as 
the effective theory of the T-dual D1-brane moving in the large
$\hT^4$.
Then, the system of $k$ D9 and $N$ D5 branes
emerged above must be identified as
T-dual to the original system of $k$ D9 and $N$ D5 branes.
Therefore, what we have seen shows that the
the T-duality of D9-D5 brane system (or other (physical) D$(p+4)$-D$p$
brane system with $p=0,1,2,3,4$) is indeed given by the
Nahm transform of instantons on $T^4$ and $\hT^4$.

One important property of T-duality is that
if the T-duality is operated twice we get back in the same
string background.
It is indeed known
as the inversion theorem \cite{BvB,DK}
that the square Nahm transform is the identity,
$(E,A)\mapsto (\hat{E},\hat{A})\mapsto (\hat{\hat{E}},\hat{\hat{A}})
\cong (E,A)$
(see also Appendix A for the expression of the isomorphism
$(\hat{\hat{E}},\hat{\hat{A}})
\cong (E,A)$).

Use of brane probe to study space-time geometry
was initiated in \cite{Douglas} and has been an important
method in string theory.
The paper \cite{Douglas} considers D9-D5 system in flat non-compact
space-time probed by a D1-brane and obtains ADHM construction of
instantons on $\R^4$
(which is nothing but the S-dual of the heterotic worldsheet
theory in \cite{WADHM}).
Since Nahm-transform is in a sense the ADHM construction for
four-torus, what we have done can be considered as
a generalization of \cite{Douglas}.
Indeed, similar argument had been used in \cite{Diaconescu}
where Nahm equation for monopoles on $\R^3$ was considered.\footnote{
Certain monopoles can be considered as instantons on
$S^1\times \R^3$ and \cite{Diaconescu}
is closely related to T-duality on $S^1$.
Nahm transform on $S^1$ (and $T^2$)
is also discussed in \cite{SethiKapu,Tsimpis} more completely than
\cite{Diaconescu}
but their argument is not applicable to T-duality on $T^4$.}
Our argument is also close to the one in \cite{SYZ}
where D3-branes wrapped on three-torus in a Calabi-Yau three-fold
are identified with D0-branes at points in the mirror.

\subsection{The General Case ---
Emergence of Brane-Anti-Brane System}

In the above discussion, we have assumed that the Dirac operator
$\Dsl=\gamma^{\mu}(\partial_{\mu}+A_{\mu}-ia_{\mu})$
has no positive-chirality zero mode
at any value of $a_{\mu}$.
In general, however, we encounter the cases where there are
positive-chirality zero modes as well.
This includes the simplest and important example of
$k=0$ where the connection $A$ is flat,
$A_{\mu}={\rm diag}(ia^1_{\mu},\ldots,ia^N_{\mu})$, in which
there are both positive and negative
chirality zero modes at $a_{\mu}=a^j_{\mu}$.
In such a case, the T-dualized system we obtain
is not a vector bundle on $\hT^4$
supported by D9-branes, but something else which we now
describe.

\newcommand{\laq}{\mbox{\large $q$}}
\newcommand{\llambda}{\mbox{\large $\lambda$}}

\subsection*{\sl Kaluza-Klein Modes and Interpretation}

Let us look at the lagrangian for the full Kaluza-Klein
modes from the hypermultiplet of 5-9 strings.
The scalar fields $Q^{\sigma}$ can be considered as 1+1
dimensional scalar fields $\laq^{\sigma}$
with values in the
infinite dimensional vector space $\Gamma(E)$ of sections of $E$.
Likewise, the fermion $\Psi$ can be considered to consist
of 1+1 dimensional spinor fields $\llambda_+$
and $\llambda_-$ of positive and negative chirality respectively
which take values in the infinite dimensional vector spaces
$\Gamma(S^+\otimes E)$ and $\Gamma(S^-\otimes E)$.
It is appropriate to consider these vector spaces of sections
of $E$, $S^+\otimes E$ and $S^-\otimes E$
as infinite-dimensional vector bundles
${\cal E}$, ${\cal E}^+$ and ${\cal E}^-$ over $\hT^4$
provided with hermitian metrics and natural connections
which we denote by ${\cal A}$, ${\cal A}_+$ and ${\cal A}_-$
respectively:
Recall that the $U(1)$ gauge symmetry of the probe D5-brane
identifies a section $\psi(x)$ at $a\in (\R^4)^*$
and a section $\e^{i\hat{n}x}\psi(x)$ at $a+\hat{n}\in(\R^4)^*$
of any of the bundles $E$, $S^+\otimes E$ and $S^-\otimes E$.
This defines the bundles ${\cal E}$, ${\cal E}^+$
and ${\cal E}^-$ over $\hT^4$.
Since this identification is unitary and does not involve $a_{\mu}$
explicitly, the hermitian products and the trivial connections
of the trivial bundles over $(\R^4)^*$
descends to hermitian products and connections of the bundles
over $\hT^4$.

Note that the components
$D^+:\Gamma(S^+\otimes E)\to\Gamma(S^-\otimes E)$ and
$D^-:\Gamma(S^-\otimes E)\to\Gamma(S^+\otimes E)$
of the Dirac operator $\Dsl$  are conjugate to
each other and that the Weizenbock formula shows 
$D^+D^-=-D^{\mu}D_{\mu}$.
Then, denoting the operator $D^+$ at $a$ as
${\cal D}(a):{\cal E}^+|_a\to{\cal E}^-|_a$,
the lagrangian can be written as
\beqa
&&\sum_{\sigma=1,2}\sum_{u=0,9}\mp
\Bigl|\Bigl(\partial_u+\partial_ua_{\mu}{\cal
    A}^{\mu}(a)-ia_u\Bigr)\laq^{\sigma}
\Bigr|^2
+\sum_{\sigma=1,2}\overline{\laq^{\sigma}}
{\cal D}(a){\cal D}(a)^{\dag}\laq^{\sigma}\nonumber\\
&&\!\!+~\overline{\llambda_+}
\Bigl(\partial_-+
\partial_-a_{\mu}{\cal A}_+^{\mu}(a)-ia_-\Bigr)\llambda_+
\,+\,
\overline{\llambda_-}
\Bigl(\partial_++
\partial_+a_{\mu}{\cal A}_-^{\mu}(a)-ia_+\Bigr)\llambda_-
\nonumber\\[0.1cm]
&&\!\!+~
\overline{\llambda_-}{\cal D}(a)\llambda_+
+\overline{\llambda_+}{\cal D}(a)^{\dag}\llambda_-.
\label{KKL}
\eeqa
We interpret (\ref{KKL}) as the lagrangian for a D1-brane probing
some bound state of infinitely many D9-brane and anti-D9-branes
wrapped on $\hT^4$ of the T-dualized system.
This is based on the following consideration.

If there are $n$ D9-branes and a D1-brane,
the 1-9 strings creates a positive-chirality fermion
on the 1+1 dimensional worldvolume of the D1-brane
which is charged
under the bifundamental representation of $U(n)\times U(1)$.
Similarly, if there are $\bar n$ anti-D9-brane and a D1-brane,
the 1-$\bar 9$ strings create a negative chirality fermion
in the bifundamental of $U(\bar n)\times U(1)$.
If there are $n$ D9-brane and $\bar n$ anti-D9-branes,
the analysis of open strings shows that,
in addition to the $U(n)\times U(\bar n)$ gauge fields
created by 9-9 and $\bar 9$-$\bar 9$ strings,
there are
tachyonic modes created by 9-$\bar 9$ strings
which are charged under $U(n)\times U(\bar n)$
as the bifundamentals $({\bf n},{\bf \bar n}^*)$
and $({\bf \bar n},{\bf n}^*)$.
(In this paper, we denote the dual of a representation $V$ by
$V^*$.)
Therefore,
the configuration without the tachyon expectation value is
unstable and is expected to
roll down toward a minimum of the tachyon potential.
Since we do not have a correct decription of some
stable bound state of a brane-anti-brane system,
we do not precisely know what happends for the theory on a D1-brane
probing such a system.
However, we may expect that the low-lying spectrum is not
very much different from the simple superposition
of those for the 1-9 system and those for the 1-$\bar 9$ system. 
In particular, there will be positive and negative chirality fermions
in $({\bf n},{\bf 1},-1)$ and $({\bf 1},{\bf \bar n},-1)$
of $U(n)\times U(\bar n)\times U(1)$ respectively (and their duals).
We may also expect that
the tachyon vev will provide a mass term in the theory of
D1-brane which couples the
positive and negative chirality fermions created by the
1-9 and 1-$\bar 9$ strings.
If it is the case, when the tachyon vev is non-zero
the coupled fermions are irrelevant in
the infra-red.
This is consistent with the expectation that
a D-brane and an anti-D-brane will annihilate via the
tachyon condensation.

Now, part of the interpretation of (\ref{KKL}) is clear.
The fermion $\llambda_+$ is interpreted as the collection of
fermions created by
the 1-9 strings in the $n\to\infty$ limit,
whereas $\llambda_-$ is interpreted as those
created by 1-$\bar 9$ strings in the $\bar n\to\infty$ limit.
The connections ${\cal A}_+$ of 
${\cal E}^+$
and ${\cal A}_-$ of ${\cal E}^-$ are the connections
of the Chan-Paton bundles supported by the D9 and anti-D9-branes
respectively.
The operators ${\cal D}:{\cal E}^+\to{\cal E}^-$
and ${\cal D}^{\dag}:{\cal E}^-\to{\cal E}^+$
are interpreted as the tachyon fields.

However, it is not obvious how to interpret the scalar fields
$\laq^{\sigma}$ which become massless at the locus where
${\cal D}(a)^{\dag}$ has a zero.
Their charge under the symmetry of the system suggest
that they are created from the 1-9 or 1-$\bar 9$ strings, but
the standard analysis shows
that the NS sector of the 1-9 or 1-$\bar 9$ strings
has the lowest mass squared $={1\over 2}$ and therefore
the corresponding bosons can never become massless.
We leave this as an open problem.
In this paper, we consider the existence of $\laq^{\sigma}$ as
a consequence of
the (0,4) supersymmetry (required from
the supersymmetry of the bound state)
which postulates the existence of the superpartner of the
fermions of mass-squared matrix
${\cal D}(a){\cal D}(a)^{\dag}$.

\subsection*{\sl Localized Degrees of Freedom}

Irrespective of the interpretation as the bound state of D9-branes
and anti-D9-branes,
it is clear that
the modes with non-zero eigenvalues of the Laplace/Dirac operators
--- ${\cal D}(a){\cal D}(a)^{\dag}$ or
${\cal D}(a)^{\dag}{\cal D}(a)$ --- are irrelevant in the
infra-red limit and can be ignored.
When ${\cal D}(a)^{\dag}$ has no kernel everywhere on $\hT^4$,
we only have to take into account the kernel of ${\cal D}(a)$,
and we are back in the cases considered in section 2.1: we obtain
the Nahm transform $\hat{E}$ supported by $k$ D9-branes
of the T-dualized system.

Something special happens when ${\cal D}(a)^{\dag}$
has a non-trivial kernel (and the kernel of ${\cal D}(a)$ jumps)
at some locus $M$ in $\hT^4$.
The mass of some supermultiplet
(two complex scalars and two Dirac fermions)
goes down toward $M$ and vanish at $M$.
In such a case, this multiplet
can no longer be ignored at least in a neighborhood of $M$.
In other words
there are some degrees of freedom localied at $M\subset \hT^4$.

We examine what this is in our favorite example of $k=0$ and flat
$A_{\mu}$. For simplicity we consider the case $N=1$.
In this case, $E$ is the trivial complex line bundle
over $T^4$ and the flat
connection is given by $A_{\mu}=ia^0_{\mu}$ (constant).
The Dirac operators are
\beq
{\cal D}(a)=
\overline{\sigma}^{\mu}(\partial_{\mu}+i(a^0_{\mu}-a_{\mu})),~~
{\cal D}(a)^{\dag}
=\sigma^{\mu}(\partial_{\mu}+i(a_{\mu}^0-a_{\mu})),
\eeq
where $\overline{\sigma}^{\mu}:S^+\to S^-$ and
$\sigma^{\mu}:S^-\to S^+$ are represented as
\beq
\sigma^1=
\left(\begin{array}{cc}
0&i\\
i&0
\end{array}\right),~~
\sigma^2=
\left(\begin{array}{cc}
0&1\\
-1&0
\end{array}\right),~~
\sigma^3=
\left(\begin{array}{cc}
i&0\\
0&-i
\end{array}\right),~~
\sigma^4=
\left(\begin{array}{cc}
1&0\\
0&1
\end{array}\right),
\label{sigmat}
\eeq
and $\overline{\sigma}^{\mu}=\sigma^{\mu\dag}$
under trivializations $S^+\cong T^4\times \C^2$ and
$S^-\cong T^4\times \C^2$.
There is no kernel for both ${\cal D}(a)$ and ${\cal D}(a)^{\dag}$
everywhere except $a=a^0$. At $a=a^0$, the constant sections
of $S^+\cong T^4\times \C^2$ and
$S^-\cong T^4\times \C^2$ become the kernels of
${\cal D}(a)$ and ${\cal D}(a)^{\dag}$ respectively
and there are nothing else.
Thus, we only have to look at the spaces
${\cal E}^+_0$ and ${\cal E}^-_0$ of these constant sections
in the vicinity of $a=a^0$.
The operator ${\cal D}(a)$ sends ${\cal E}_0^+$ to ${\cal E}_0^-$
and behaves as
\beq
{\cal D}(a)=-i\overline{\sigma}^{\mu}(a-a^0)_{\mu}.
\label{Dat}
\eeq
What these all mean in the 1+1 dimensional theory of the probe
is that there are two complex bosons
$q_0^{\sigma}$ and two positive and negative chirality fermions
$\lambda_+^{\alpha}$, $\lambda_-^{\dot{\alpha}}$
which are localized near $a=a^0$.
Their mass squared behaves as
${\cal D}(a){\cal D}(a)^{\dag}=|a-a^0|^2$.
Note that $q_0^{\sigma}$ transforms in the positive
spinor representation of $SO(4)$ in the $x^{5678}$ directions
while $\lambda_+^{\alpha}$ and $\lambda_-^{\dot{\alpha}}$
transforms in the negative and positive spinor representations
of $SO(4)$ in the $\hx_{1234}$ directions.
These are clearly the properties of the 1-5 string 
in the system of D1-brane probing the D5-brane located at $a=a^0$.
Thus, this localized degrees of freedom can be identified as
the D5-brane at $a^0\in \hT^4$ of the T-dualized system.
This is of course what is expected from the known relation of
the Wilson line
and the position of the D-branes under T-duality \cite{DLP}.

The same conclusion can be deduced also
in the picture of D9-anti-D9-brane system.
Although there are infinitely many D9 and anti-D9-branes,
we may regularize the system by discarding infinite high
level modes which are irrelevant in the infra-red limit.
One obvious choice appears to be the one to discard everything but
the constant modes.
However, there is a technical difficulty. Because of the
identification of the sections $\psi(x)$ at $a$ and
$\e^{i\hat{n}x}\psi(x)$ at $a+\hat{n}$,
the space of constant sections (defined as ${\cal E}_0^{\pm}$
in a neighborhood of $a^0$)
does not extend as a globally
defined finite-dimensional vector bundle over $\hT^4$.
Instead of trying to find a finite-dimensional subbundle
of ${\cal E}^{\pm}$, we can consider a more wild but reasonable
regularization.
Since the only relevant information is the behaviour
of the lowest modes in the vicinity of
$a=a^0$,
we can approximate the infinite-dimensional
bundles ${\cal E}^{\pm}$ with the tachyon
${\cal D}:{\cal E}^+\to {\cal E}^-$
by rank two vector bundles $\widetilde{\cal E}_0^{\pm}$ with
a tachyon $\widetilde{\cal D}:\widetilde{\cal E}_0^{+}\to
\widetilde{\cal E}_0^{-}$ defined globally over $\hT^4$
such that
$\widetilde{\cal D}:\widetilde{\cal E}_0^{+}\to
\widetilde{\cal E}_0^{-}$
is the same as ${\cal D}:{\cal E}_0^+\to{\cal E}_0^-$
when restricted to a neighborhood of $a=a^0$
and is an isomorphism outside.
Since ${\cal D}(a):{\cal E}_0^+\to{\cal E}_0^-$ 
given by (\ref{Dat})
has a winding number one on the three-sphere surrounding $a=a^0$
and $\widetilde{\cal D}(a)$
is an isomorphism outside,
the instanton numbers of
$\widetilde{\cal E}_0^{+}$ and
$\widetilde{\cal E}_0^{-}$ must differ by one,
irrespective of the choice of the extension
$\widetilde{\cal E}_0^{\pm}$
of ${\cal E}_0^{\pm}$.
This together with the Chern-Simons coupling
on the (anti-)D-brane
shows that there is a single
D5-brane.
Since the degrees of freedom is localized at $a=a^0$,
the D5-brane must be at $a^0\in\hT^4$.

\subsection*{\sl Mukai's Fourier Transform}

We have described what we obtain as the T-dualized system in the general
case, using the effective theory on the probe.
We have also gave an interpretation as some bound state of
D9-branes and anti-D9-branes.
However, we have not described it in a mathematical language
except for the generic case where the kernel of ${\cal D}(a)^{\dag}$
is constantly zero and the T-duality is described as the Nahm transform.
In algebraic geometry, there is a transform called Mukai's Fourier
transform \cite{Mu} which can be considered as a generalization of
the Nahm transform.
It is a transform of an object in a category called
derived category of sheaves \cite{Residues} on an abelian variety $X$
(a complex torus embedded in a projective space)
denoted by ${\bf D}(X)$ to an object of ${\bf D}(\widehat{X})$
where $\widehat{X}$ is the dual torus of $X$
(which is again an abelian variety).
The category ${\bf D}(X)$ includes as its objects
holomorphic vector bundles on 
$X$, and for the case where $X$ is four-torus
and in a region where there is a one to one correspondence between
holomorphic bundles and (anti-)self-dual connections,
Mukai's Fourier transform agrees with the Nahm transform.
This suggests that the correct mathematical language
to describe what we have obtained should not be far from
the derived category of sheaves, at least in the case where
$T^4$ has a structure of an abelian variety.
Indeed, a general object of ${\bf D}(X)$ is a complex of sheaves
which is reminiscent of our Dirac complex, and also, sheaves can localize
on a subvariety of $X$.
It is an interesting problem to find the precise relation
(though we do not attempt to solve here).

\subsection*{\sl The Inverse Transform}

We have started from a connection of a vector bundle on $T^4$
supported by D9-branes and obtained, as its T-dual image,
an object which is
something 
more general than a connection of a vector bundle,
presumably supported by D9 and anti-D9-branes.
Then, a natural question is what happens when
T-duality is applied to such an object.
To test the interpretation as the D9-anti-D9 bound state,
assuming some property of branes probing such a system,
we consider the T-duality of the object which is obtained as
the T-dual image of our favorite example of $k=0$, $N=1$ flat
connection $A_{\mu}=ia^0_{\mu}$.
We must get back the original flat connection.
We will work in the finite-dimensional approximation of
${\cal D}:{\cal E}^+\to{\cal E}^-$ introduced before and will
focus our attention to a neigorhood of $a=a^0$,
ignoring the global issue on $\hT^4$
which yields only a subleading modification in the
present discussion.

Thus, our starting point is the two pairs of D9 and anti-D9-branes
wrapped on $\hT^4$ supporting the rank-two Chan-Paton bundles
 $E^+$ and $E^-$
and the tachyon field $T:E^+\to E^-$
which behaves near $a=a^0$ as (\ref{Dat}),
$T(a)=-i\overline{\sigma}^{\mu}(a-a^0)_{\mu}$.
We probe this system by a D5-brane wrapped on $\hT^4$.
As in the case for a D1-brane probing the brane-anti-brane system, we
assume that the low-lying fermionic spectrum on the D5-brane
is the same as the superposition of those for the D5-D9 system and
the D5-anti-D9 system. Also, we assume that the tachyon expectation
value yields the fermion mass term.
Thus, the probe theory has 5+1 dimensional negative
and positive chirality fermions with values in
$E^+$ and $E^-$ which we denote by $\Psi_-$ and $\Psi_+$ respectively.
The lagrangian for these fields is
\beq
\overline{\Psi}_-\Dsl^-\Psi_-
+\overline{\Psi}_+\Dsl^+\Psi_+
+\overline{\Psi}_+T\Psi_-
+\overline{\Psi}_-T^{\dag}\Psi_+,
\eeq
where $\Dsl^{\mp}$ are the 5+1 dimensional Dirac operator
associated with the covariant derivatives
$D_{\pm}=\partial_{\pm}-ia_{\pm}$ and
\beq
D^{\mu}={\partial\over\partial a_{\mu}}-ix^{\mu}
\eeq
where $x^{\mu}$ stands for the
flat $U(1)$ gauge field on $\hT^4$. Here we have ignored the
gauge fields of the bundles $E^{\pm}$.

Massless fermions in the reduced 1+1 dimensions
would come from the solutions of the
equations
of motion on the $\hT^4$ factor
\beq
\left(\begin{array}{cc}
T&\hat{D}^-\\
\hat{D}^+&T^{\dag}
\end{array}\right)
\left(\begin{array}{c}
\psi_{++}\\
\psi_{--}
\end{array}\right)=0,~~
\left(\begin{array}{cc}
T&\hat{D}^+\\
\hat{D}^-&T^{\dag}
\end{array}\right)
\left(\begin{array}{c}
\psi_{-+}\\
\psi_{+-}
\end{array}\right)=0,
\label{zeromeq}
\eeq
for $\psi_{++}\in \Gamma(\hat{S}^+\otimes E^+)$,
$\psi_{-+}\in\Gamma(\hat{S}^-\otimes E^+)$ etc,
where $\hat{S}^{\pm}$ are negative and postitive
spinor bundles on $\hT^4$ and $\hat{D}^{\pm}$ are the components
of the Dirac operator associated with $D^{\mu}$.
A solution to the equation on the left (resp. right)
would lead to a positive (resp. negative) chirality fermion in
1+1 dimensions.

Now let us take a closer look at these equations.
We first recall some facts about the spinor representations
in four-dimensions.
Let ${\bf 2}_{\pm}$ be the spinor representation of
$Spin(4)=SU(2)_+\times SU(2)_-$ which are
sent to each other by the gamma matrices
$\overline{\sigma}^{\mu}:{\bf 2}_+\to{\bf 2}_-$ and
$\sigma^{\mu}:{\bf 2}_-\to{\bf 2}_+$.
We note that there are isomorphisms $\epsilon_{\pm}:
{\bf 2}_{\pm}\to{\bf 2}_{\pm}^*$ of $SU(2)_{\pm}$ representations
such that
$\epsilon_-\overline{\sigma}^{\mu}={}^t\sigma^{\mu}\epsilon_+$.
Now, 
using the metric we identify the tangent spaces of $T^4$ and $\hT^4$,
and thus
we can consider both
$S^{\pm}$ and $\hat{S}^{\pm}$
as the trivial bundles with the common fibre ${\bf 2}_{\pm}$.
In particular, the Dirac operators on $\hT^4$
can be respresented as
$\hat{D}^+=\overline{\sigma}^{\mu}D^{\mu}$
and
$\hat{D}^-=\sigma^{\mu}D^{\mu}$.
On the other hand, the bundles $E^{\pm}$
look like the bundles with fibres ${\bf 2}_{\pm}$ in the vicinity of
$a=a^0$ where the tachyon is given by
$T(a)=-i\overline{\sigma}^{\mu}(a-a^0)_{\mu}$.
For convenience, using the isomorphisms
$\epsilon_{\pm}$,
 we shall consider $E^{\pm}$
to have fibres ${\bf 2}_{\pm}^*$
and the tachyon is represented as
\beq
T(a)=-i\,{}^t\!\sigma^{\mu}(a-a^0)_{\mu}.
\eeq
Then, we can consider
$\psi_{++}$, $\psi_{--}$, $\psi_{-+}$ and $\psi_{+-}$
as functions taking values in ${\bf 2}_+\otimes{\bf 2}_+^*$,
${\bf 2}_-\otimes{\bf 2}_-^*$,
${\bf 2}_-\otimes{\bf 2}_+^*$ and
${\bf 2}_+\otimes{\bf 2}_-^*$ respectively, and the equations
(\ref{zeromeq}) look like
\beq
\left\{\begin{array}{l}
\left({\partial\over\partial a_{\mu}}-ix^{\mu}\right)
\sigma^{\mu}\psi_{--}
=i(a-a^0)_{\mu}\psi_{++}\sigma^{\mu},\\[0.1cm]
\left({\partial\over\partial a_{\mu}}-ix^{\mu}\right)
\overline{\sigma}^{\mu}\psi_{++}
=-i(a-a^0)_{\mu}\psi_{--}\overline{\sigma}^{\mu},
\end{array}\right.
~
\left\{\begin{array}{l}
\left({\partial\over\partial a_{\mu}}-ix^{\mu}\right)
\overline{\sigma}^{\mu}\psi_{+-}
=i(a-a^0)_{\mu}\psi_{-+}\sigma^{\mu},\\[0.1cm]
\left({\partial\over\partial a_{\mu}}-ix^{\mu}\right)
\sigma^{\mu}\psi_{-+}
=-i(a-a^0)_{\mu}\psi_{+-}\overline{\sigma}^{\mu}.
\end{array}\right.
\eeq
The equations on the left are solved by
\beqa
\psi_{++}&=&1_{{\bf 2}_+}
\exp\left(-{1\over 2}|a-a^0|^2+ixa\right),\\
\psi_{--}&=&i\,1_{{\bf 2}_-}
\exp\left(-{1\over 2}|a-a^0|^2+ixa\right),
\eeqa
while the equations on the right have no solution.
This is true for any values of $x^{\mu}\in \R^4$.
Note that
this solution is single valued as a function of $x_{\mu}\in T^4$
since it obeys the correct
periodicity
with respect to
$x^{\mu}\to x^{\mu}+n^{\mu}$ ($n\in\Lambda$):
\beq
\psi(a) \,\mbox{at $x$}\,\equiv
\e^{ina}\psi(a) \,\mbox{at $x+n$}
\eeq
which is dictated by the $U(1)$ gauge symmetry of the probe
D5-brane.

Thus, we have a single positive-chirality fermion
in the effective 1+1 dimensional theory.
This is interpreted as the 1-9 string mode of the D1-brane probing a
D9-brane.
Since the above solution for $(\psi_{++},\psi_{--})$
is single valued and nonwhere vanishing
over the whole space $T^4$,
the Chan-Paton bundle of the D9-brane is topologically trivial.
The gauge field on this trivial bundle can be computed as
\beq
A_{\mu}(x)=\int_{\hT^4}\dd^4 a\,\,
(\psi_{++}^{\dag},\psi_{--}^{\dag}){\partial\over\partial x^{\mu}}
\left(\begin{array}{c}
\!\psi_{++}\!\\
\!\psi_{--}\!
\end{array}\right)
\,=\,ia^0_{\mu}
\eeq
when $\psi_{++}$ and $\psi_{--}$ are correctly normalized.
Thus, we have recovered the original flat connection
supported by a single D9-brane.

\section{Orientifold and Orbifold}

We apply the argument of the previous section to the case
where the four-torus is 
at an orientifold fixed plane or is modded out by $\Z_2$
orbifold action.
We describe the T-duality in terms of the gauge field configuration
representing the system.
This will lead us to find a Nahm transform for instantons on $T^4$ 
with orthogonal/symplectic gauge groups or on orbifold
$T^4/\Z_2$.
Here we only consider the generic case where the transformed object
is a vector bundle, and more general case will not be presented
since that would be a repetition of section 2.2.
(However, the general case will be included in section 4.)

\subsection{D-branes Wrapped on
$\Z_2$ Orientifold/Orbifold of Four-Torus}

As preliminaries, we provide the gauge theory description
of D-branes wrapped on $\Z_2$ orientifold/orbifold of four-torus
where $\Z_2$ acts on the torus via the inversion --- the sign flip
of all four flat coordinates.
(For orientifold, we consider here the case where
D-branes and orientifold-planes are parallel except
the four-torus directions.)
We apply the method of \cite{Taylor} to find the description.
Similar analysis for orientifold
has been done in \cite{BMot,Hora} and most close
one is in \cite{Rey}, while the orbifold case was analyzed
in \cite{RW,GLY}.

\subsection*{(i) Orientifold of $SO$-Type}

We first consider the case of orientifold of $SO$-type.
Thus, we would like to find a gauge theory description of, say,
$2k$ D4-branes in Type IIA orientifold on
$\R\times (\hT^4\times \R^5)/\Z_2$ which are wrapped on the $\hT^4$
directions.
For this, we start with the T-dualized system
of $2k$ D0-branes in Type IIA orientifold on
$\R\times T^4\times \R^5/\Z_2$ and perform the Fourier transform as in
\cite{Taylor}.

If $T^4$ were $\R^4$, the theory on the D0-branes would be
the supersymmetric quantum mechanics with eight supercharges
which can be obtained as the dimensional reduction of four-dimensional
$N=2$ $Sp(k)$ gauge theory with a hypermultiplet in the anti-symmetric
representation. The bosonic fields in such a theory is
$Sp(k)$ gauge field $A_0$ (which can be gauged away),
the scalars $X^p$ ($p=5,6,7,8,9$) in the adjoint representation,
and the scalars $X^{\mu}$ ($\mu=1,2,3,4$) in the anti-symmetric
representation. The adjoint and anti-symmetric representation fields,
$X^p$ and $X^{\mu}$,
can be represented by $2k\times 2k$ anti-hermitian and hermitian
matrices respectively, both of which obey
\beq
J\overline{X}=XJ,~~~
J=
\left(\begin{array}{cc}
0&\!-{\bf 1}_k\\
\,{\bf 1}_k\!&\!0
\end{array}\right),
\label{Jrel}
\eeq
where $\overline{X}$ is the complex
conjugation of $X$ and
 ${\bf 1}_k$ is the $k\times k$ unit matix.

For $T^4=\R^4/\Lambda$,
we must take into accout the open strings winding around 1-cycles
(and ending on the D0-branes).
Thus, as in \cite{Taylor}, we must replace the $Sp(k)$
gauge group to the symplectic group ``$Sp(k|\Lambda|)$''
of infinite rank.
The bosonic fields in the theory can be represented by
$2k\times 2k$ matrices $X^p_{nm}$, $X^{\mu}_{nm}$ parametrized by
$(n,m)\in \Lambda\times \Lambda$ which obey
$X^{p\dag}_{nm}+X^p_{mn}=0$, $X^{\mu\dag}_{nm}=X^{\mu}_{mn}$
and also (\ref{Jrel}) for each $X=X^p_{nm}, X^{\mu}_{nm}$.
There are also periodicity conditions
$X^p_{(n+n^{\prime})(m+n^{\prime})}=X^p_{nm}$
and $X^{\mu}_{(n+n^{\prime})(m+n^{\prime})}=
X^{\mu}_{nm}+(n^{\prime})^{\mu}\delta_{n,m}{\bf 1}_{2k}$.
Let us put
\beqa
X^p(\hx)&=&\sum_{n\in \Lambda}\e^{in\hx}X_{n0}^p,\\
A^{\mu}(\hx)&=&i\sum_{n\in\Lambda}\e^{in\hx}X^{\mu}_{n0},
\eeqa
where $\hx=(\hx_{\mu})$ are coordinates of the dual
torus $\hT^4=(\R^4)^*/2\pi\Lambda^*$.
Then, these are functions on $\hT^4$
with values in $2k\times 2k$
anti-hermitian matrices and satisfy
\beqa
&&J\overline{X^p(\hx)}=X^p(-\hx)J,\label{SOcond1}\\
&&J\overline{A^{\mu}(\hx)}=-A^{\mu}(-\hx)J.\label{SOcond2}
\eeqa
The $Sp(k|\Lambda|)$ gauge transformation is represented as the
transformation of the fields $X^p(\hx), A^{\mu}(\hx)$
given infinitesimally by
$\delta X^p(\hx)=[X^p(\hx),\alpha(\hx)]$ and
$\delta A^{\mu}(\hx)=[A^{\mu}(\hx),\alpha(\hx)]
+\partial^{\mu}\alpha(\hx)$ where $\alpha(\hx)$ is a function on
$\hT^4$ which satisfy the same conditions as $X^p(\hx)$.
The lagrangian is the same as the one for the $4+1$ dimensional
$N=2$ supersymmetric (sixteen supercharges)
$U(2k)$ Yang-Mills theory formulated on $\R\times \hT^4$.
The number of unbroken
supersymmetries are reduced to eight by the constraints
(\ref{SOcond1}) and (\ref{SOcond2}) on the fields.

This is the gauge theory description of
wrapped D4 branes in the Type IIA orthogonal-orientifold
on $\R\times (\hT^4\times \R^5)/\Z_2$.
It is obvious how to generalize this to the system of wrapped
D$(p+4)$-branes in the Type II orientifold on
$\R^{p+1}\times (\hT^4\times \R^{5-p})/\Z_2$
with O$p$-plane of $SO$-type at each of the sixteen fixed point.

The conditions (\ref{SOcond1}) and (\ref{SOcond2}) can be
restated in a way
which applies also to the case where the gauge bundle on
$\hT^4$ is topologically non-trivial.
Let $E$ be a $U(2k)$ bundle over $\hT^4$
(a rank $2k$ complex vector bundle
provided with a hermitian fibre metric $(\,,\,)$).
The fields $X^p(\hx)$ and $A^{\mu}(\hx)$ represent respectively a
section $X^p$ of the adjoint bundle of $E$
(an anti-hermitian endomorphism of $E$)
and a unitary connection $\nabla$ of $E$
(a connection preserving the hermitian metric).
Now suppose we have a family of anti-liner maps
$J=(J_{\hx})$
sending the fibre at $\hx$ to the fibre at
$-\hx$:
\beq
J_{\hx}:E_{\hx}\longrightarrow E_{-\hx},~~~\mbox{such that~
$J_{-\hx}J_{\hx}=-{\rm id}_{\hx}$}
\eeq
which is isometric in the sense that
$(J_{\hx}v,J_{\hx}w)=(w,v)$.
We shall call such a family $J$
a {\it symplectic structure of $E$ over the inversion}
$\hx\mapsto -\hx$ of $\hT^4$, and such a pair $(E,J)$
a {\it symplectic orientibundle over
the (orthogonal) orientifold} ~$\hT^4/\Z_2$.
Such anti-linear maps $J$ can be locally represented by the matrix
$(J_{ji})$ in (\ref{Jrel}) as
$J_{\hx}e_i(\hx)=e^{\prime}_j(-\hx)J_{ji}$ where 
$e_i$ is an orthonormal frame of $E$ defined in an open subset
of $\hT^4$
and $e^{\prime}_j$ is another one defined in the inversion image
(we shall call such a pair of frames a {\it symplectic frame}).
Then, the condition (\ref{SOcond1}) means that $X^p$
commutes with $J$, $J_{\hx}X^p(\hx)=X^p(-\hx)J_{\hx}$,
and the condition (\ref{SOcond2}) requires that
the connection $\nabla$
preserves the symplectic structure $J$ in the sense that
$J\nabla_{\!X}s=\nabla_{\!(-X)}Js$
for a (local) section $s$ and a vector field $X$ on $\hT^4$
where $Js$ is another section defined by
$J_{\hx}s(\hx)=(Js)(-\hx)$ and $-X$ is the image of $X$ under the
inversion $\hx\mapsto -\hx$.
The gauge transformations are (local) unitary automorphisms of $E$
which commute with the map $J$.
\footnote{
There is another way to state the conditions:
Let us define 
a bilinear pairing
$\la\,,\,\ra$ of the fibres at $\hx$ and $-\hx$ by
$\la v_{\hx},w_{-\hx}\ra=(Jv_{\hx},w_{-\hx})$. Then, this is a
non-degenerate skew-symmetric form (skew-symmetric in the sense 
that $\la v_{\hx},w_{-\hx}\ra=-\la w_{-\hx},v_{\hx}\ra$).
Then, the conditions (\ref{SOcond1}) and (\ref{SOcond2}) means that
the gauge transformations and
the connection $\nabla$ should preserve the skew-form $\la\,,\,\ra$.}

One can show that a $U(2k)$ bundle on $\hT^4$ with arbitrary
instanton number $N$ admits a symplectic structure over the inversion.
Let us choose a neighborhood $D$ of $\hx=0$ given by $|\hx|\leq
\epsilon$ and let $\overline{\hT^4\setminus D}$ be the
outside region.
A $U(2k)$ bundle $E$ of instanton number $N$ can be
constructed by glueing the trivial bundles over $D$ and
$\overline{\hT^4\setminus D}$
at the boundary $S^3=\{|\hx|=\epsilon\}$
by a transition function $g:S^3\to U(2k)$ of winding number $N$.
We define the anti-linear map
$J_{\hx}:E_{\hx}\to E_{-\hx}$ over $\overline{\hT^4\setminus D}$
as the complex conjugation followed by the multiplication
by the matrix $J$ in (\ref{Jrel})
with respect to the trivialization we started with.
At the boundary $S^3$,
the matrix to be multiplied is expressed as
$J(\hx)=g(-\hx)^{-1}J\overline{g(\hx)}$
with respect to the trivialization that extends over $D$.
We can extend $J_{\hx}$ to the interior of $D$
if $J(\hx)$ can be extended to a function on $D$ satisfying
$J(-\hx)\overline{J(\hx)}=-1$ and
$J(\hx)^{\dag}J(\hx)=1$.
Let us embed $Sp(1)$ in $Sp(k)\subset U(2k)$
via $h\in Sp(1)\mapsto {\rm diag}(h,1,\ldots,1)\in Sp(k)$.
Then, a map $\hx\in S^3=Sp(1)\mapsto \hx^N \in Sp(1)$
induces a map $g: S^3\to U(2k)$ of winding number $N$
which yields $J(\hx)=(-1)^NJ$ on $S^3$.
Thus, $J_{\hx}$ extends to the interior of $D$ and hence to
all over $\hT^4$, defining a symplectic structure over the inversion.

\subsection*{(ii) Orientifold of $Sp$-Type}

We next consider the case of orientifold of $Sp$-type.
We shall find a gauge theory description of
$k$ D4-branes in Type IIA orientifold on
$\R\times (\hT^4\times \R^5)/\Z_2$ which are wrapped on the $\hT^4$
directions.
We start with the T-dualized system
of $k$ D0-branes in Type IIA orientifold on
$\R\times T^4\times \R^5/\Z_2$ and
proceed as in the previous case.
The differece is that
(in the case where $T^4$ is replaced by $\R^4$)
the gauge group is now $O(k)$ and the hypermultiplet is in the
second rank symmetric tensor representation.
The effect is to eliminate the matrix $J$ in every formula in
the previous case.

Thus, the theory of D4-branes wrapped on $\hT^4/\Z_2$ orientifold
contains bosonic fields
$X^p(\hx)$ ($p=5,6,7,8,9$) and $A^{\mu}(\hx)$ ($\mu=1,2,3,4$)
with values in $k\times k$
anti-hermitian matrices obeying
\beqa
&&\overline{X^p(\hx)}=X^p(-\hx),\label{Spcond1}\\
&&\overline{A^{\mu}(\hx)}=-A^{\mu}(-\hx).\label{Spcond2}
\eeqa
The infinitesimal gauge transformation parameter $\alpha(\hx)$
obey the same conditions as $X^p(\hx)$.
The lagrangian is the same as the one for $4+1$ dimensional
$N=2$ $U(k)$ super-Yang-Mills theory on $\R\times \hT^4$
but the supersymmetries
are reduced to eight by the constarints
on the fields.

We restate
the conditions (\ref{Spcond1}) and (\ref{Spcond2})
in a general set up.
Let $E$ be a $U(k)$ bundle over $\hT^4$.
The fields $X^p(\hx)$ and $A^{\mu}(\hx)$ represent respectively a
section $X^p$ of the adjoint bundle of $E$
and a unitary connection $\nabla$ of $E$.
Now suppose we have a family of anti-liner maps
$I=(I_{\hx})$
sending the fibre at $\hx$ to the fibre at
$-\hx$:
\beq
I_{\hx}:E_{\hx}\longrightarrow E_{-\hx},~~~\mbox{such that~
$I_{-\hx}I_{\hx}={\rm id}_{\hx}$}
\eeq
which is isometric in the sense that
$(I_{\hx}v,I_{\hx}w)=(w,v)$.
We shall call such a family $I$
an {\it orthogonal structure of $E$ over the inversion}
$\hx\to -\hx$ of $\hT^4$,
and such a pair $(E,I)$ an {\it orthogonal orientibundle over
the (symplectic) orientifold}~ $\hT^4/\Z_2$.
Such anti-linear maps $I$ can be locally represented as
$I_{\hx}e_i(\hx)=e^{\prime}_i(-\hx)$ where 
$e_i$ is an orthonormal frame of $E$ defined in an open subset
of $\hT^4$
and $e^{\prime}_j$ is another one defined in the inversion image
(we shall call such a pair of frames a {\it real frame}).
Then, the condition (\ref{Spcond1}) means that $X^p$
commutes with $I$
and the condition (\ref{Spcond2}) requires that
the connection $\nabla$
preserves the symplectic structure $I$
(in the similar sense as before).
The gauge transformations are (local) unitary automorphisms of $E$
which commute with the map $I$.
\footnote{
A bilinear pairing
$\la\,,\,\ra$ of the fibres at $\hx$ and $-\hx$ defined by
$\la v_{\hx},w_{-\hx}\ra=(Jv_{\hx},w_{-\hx})$ is a
non-degenerate symmetric form (symmetric in the sense 
that $\la v_{\hx},w_{-\hx}\ra=\la w_{-\hx},v_{\hx}\ra$).
Then, the conditions (\ref{Spcond1}) and (\ref{Spcond2}) means that
the gauge transformations and the connection $\nabla$
should preserve the form $\la\,,\,\ra$.}

One can show that a $U(k)$ bundle on $\hT^4$
admits an orthogonal structure over the inversion
provided the instanton number is even, say  $2N$.
The construction is as in the previous case.
Thus, we only have to show that the function
$I(\hx)=g(-\hx)^{-1}\overline{g(\hx)}$ on $S^3=\{|\hx|=\epsilon\}$
extends to a function defined on $|\hx|\leq\epsilon$ obeying
$I(-\hx)\overline{I(\hx)}=1$ and
$I(\hx)^{\dag}I(\hx)=1$ if $g:S^3\to U(k)$
is a map of even winding number $2N$. 
Let us embed $Sp(1)$ in $SO(k)\subset U(k)$
via $h\in Sp(1)\mapsto [(h,1)]\in (Sp(1)\times
Sp(1))/\Z_2=SO(4)\subset SO(k)$.
Then, a map $\hx\in S^3=Sp(1)\mapsto \hx^N \in Sp(1)$
induces a map $g: S^3\to U(k)$ of winding number $2N$.
This yields $I(\hx)=(-1)^N$ which obviously extends to
$|\hx|\leq\epsilon$.
Note that this construction applies only to the case of even instanton
numbers (presumably there is no orthogonal
structure over the inversion
for odd instanton numbers because of $\pi_3(U(k)/O(k))=\Z_2$).

\subsection*{(iii) Orbifold}

\newcommand{\vph}{\varphi}

Finally, we provide a gauge theory description of D-branes wrapped
on the orbifold\footnote{
Here, we consider string theory
based on
the orbifold CFT on the worldsheet.
When $T^4/\Z_2$ is considered as a singular K3-surafce,
NS-NS $B$-field has period $\pi$ for each of the sixteen
vanishing two-cycles.}
$T^4/\Z_2$.
We consider here $2N$ D4-branes (the generalization to
$p>4$ branes is obvious).

We start with Type IIA string theory on the dual $\hT^4/\Z_2$ 
with $2N$ D0-branes at points of $\hT^4$, and proceed as before
following \cite{Taylor}.
The bosonic fields in the theory of lowlying open string modes
can be represented by $2N\times 2N$ matrices
$X_{\hat{n}\hat{m}}^p, \hat{X}_{\mu,\hat{n}\hat{m}}$ parametrized by
$(\hat{n},\hat{m})\in 2\pi \Lambda^*\times 2\pi\Lambda^*$
which obey the usual conditions
$X^{p\dag}_{\hat{n}\hat{m}}+X^p_{\hat{m}\hat{n}}=0$,
$\hat{X}^{\dag}_{\mu,\hat{n}\hat{m}}=\hat{X}_{\mu,\hat{m}\hat{n}}$,
$X^p_{(\hat{n}+\hat{n}^{\prime})(\hat{m}+\hat{n}^{\prime})}
=X^p_{\hat{n}\hat{m}}$
and $\hat{X}_{\mu,(\hat{n}+\hat{n}^{\prime})(\hat{m}
+\hat{n}^{\prime})}=
\hat{X}_{\mu,\hat{n}\hat{m}}+(\hat{n}^{\prime})_{\mu}
\delta_{\hat{n},\hat{m}}{\bf 1}_{2N}$.
The $\Z_2$ orbifold projection amounts to
$\Phi X^p_{\hat{n}\hat{m}}\Phi =X^p_{-\hat{n}-\hat{m}}$,
$\Phi \hat{X}_{\mu,\hat{n}\hat{m}}\Phi =-\hat{X}_{\mu,-\hat{n}-\hat{m}}$,
where
\beq
\Phi=\left(
\begin{array}{cc}
0&{\bf 1}_N\\
{\bf 1}_N&0
\end{array}
\right).
\label{Pdef}
\eeq
Here we assumed that the $\Z_2$ orbifold action on Chan-Paton factor
is in a sum of copies of the regular representation,
but we can relax this condition so that $\Phi$ is an arbitrary unitary
matrix that squares to $1$.
Such matrices are classified by
${\rm Tr}\,\Phi$ up to similarity transformation.\footnote{
If ${\rm Tr}\,\Phi\ne 0$, there must be a fractional brane stuck at
the $\Z_2$ fixed point \cite{DoMo}.}
Let us put
\beqa
X^p(x)&=&\sum_{\hat{n}\in 2\pi\Lambda^*}\e^{i\hat{n}x}
X_{\hat{n}\hat{0}}^p,\\
A_{\mu}(x)&=&i\sum_{\hat{n}\in 2\pi\Lambda^*}\e^{i\hat{n}x}
\hat{X}_{\mu,\hat{n}\hat{0}},
\eeqa
where $x=(x^{\mu})$ are coordinates of the torus $T^4$.
Then, these are functions on $T^4$
with values in $2N\times 2N$
anti-hermitian matrices and satisfy
\beqa
&&\Phi X^p(x)\Phi =X^p(-x),\label{Pcond1}\\
&&\Phi A_{\mu}(x)\Phi =-A_{\mu}(-x).\label{Pcond2}
\eeqa
The gauge transformation is represented
infinitesimally by
$\delta X^p(x)=[X^p(x),\alpha(x)]$ and
$\delta A_{\mu}(x)=[A_{\mu}(x),\alpha(x)]
+\partial_{\mu}\alpha(x)$ where $\alpha(x)$ is a function on
$T^4$ which satisfy the same conditions as $X^p(x)$.
The lagrangian is the same as the one for the $4+1$ dimensional
$N=2$ $U(2N)$ super-Yang-Mills theory on
$\R\times T^4$, but the number of
supersymmetries are reduced to eight by the constraints
on the fields.

We restate
the conditions (\ref{Pcond1}) and (\ref{Pcond2})
in a general set up where the gauge bundle on $T^4$ is not 
necessarily topologically trivial.
Let $E$ be a $U(2k)$ bundle over $T^4$.
The fields $X^p(x)$ and $A_{\mu}(x)$ represent respectively a
section $X^p$ of the adjoint bundle
and a unitary connection $\nabla$.
Now suppose we have a family of linear maps
$\vph=(\vph_x)$
sending the fibre at $x$ to the fibre at
$-x$:
\beq
\vph_x:E_{x}\longrightarrow E_{-x},~~~\mbox{such that~
$\vph_{-x}\vph_{x}={\rm id}_{x}$}
\eeq
which is unitary in the sense that
$(\vph_{x}v,\vph_{x}w)=(v,w)$.
We shall call such a family $\vph$
a {\it lift to $E$ of the inversion}
$x\mapsto -x$ of $T^4$,
and such a pair $(E,\vph)$ an {\it orbibundle over the
orbifold $T^4/\Z_2$}.\footnote{
A more standard terminology (after the $\Z_2$ quotient)
is {\it V-bundle over V-manifold}.
The name ``orbibundle''
is due to K. Fukaya and K. Ono (as far as I know).
``Orientibundle'' which I introduced in the previous part of
the paper simply follows this terminology.}
Then, the condition (\ref{Pcond1}) means that $X^p$
commutes with $\vph$ and the condition (\ref{Pcond2}) requires that
the connection $\nabla$
preserves $\vph$ in the usual sense.
The gauge transformations are (local) unitary automorphisms of $E$
which commute with the map $\vph$.
Two lifts of the inversion cannot be equivalent
if the traces are different at any of the sixteen fixed points
$\{x_i\}$.
We shall call a lift $\vph$ {\it traceless} if ${\rm Tr}\,\vph_{x_i}=0$
at all of the fixed points.

One can show that a $U(2N)$ bundle on $T^4$
admits a traceless
lift of the inversion provided the instanton number is
even, say $2k$. As before,
we only have to show that the function
$\Phi(x)=g(-x)^{-1}\Phi g(x)$ on $S^3=\{|x|=\epsilon\}$
extends to $|x|\leq \epsilon$
as a function obeying $\Phi(-x)\Phi(x)=1$ and
$\Phi(x)^{\dag}\Phi(x)=1$,
if $g:S^3\to U(2N)$ is a map of even winding number $2k$.
We choose $g(x)$ to be ${\rm diag}(g_2(x),{\bf 1}_{N-2},
g_2(x),{\bf 1}_{N-2})$ where $g_2(x)$ is a map $S^3\to SU(2)$
of winding number $k$ such that $g_2(-x)=(-1)^kg_2(x)$
(it is easy to construct such $g_2(x)$: for example,
identify $x\in S^3$ as a unit quaternion
and put $g_2(x)=x^k \in Sp(1)=SU(2)$).
Then, $\Phi(x)$ is a constant unitary matrix $\Phi^{\prime}$
such that $\Phi^{\prime 2}={\bf 1}_{2N}$
and ${\rm Tr}\,\Phi^{\prime}=0$, and therefore extends to
$|x|\leq\epsilon$.

\subsection{Nahm Transform as T-Duality}
\newcommand{\rma}{{\rm a}}
\newcommand{\rmb}{{\rm b}}

We now describe T-duality in terms of the gauge field configuration
representing the system.
We consider T-duality transform of D$(p+4)$ and D$p$-branes
on $T^4$ where $T^4$ is
(i) at the fixed plane of orthogonal-orientifold,
(ii) at the fixed plane of symplectic-orientifold,
and (iii) divided by a $\Z_2$ orbifold action.

\subsection*{(i) Orthogonal Bundle $\leftrightarrow$
Symplectic Orientibundle}

We first consider an (unphysical) Type I string theory
(or equivalently Type IIB orientifold of $SO$-type) on
$\R^6\times T^4$ with $N$ D9-branes wrapped on $T^4$
and $2k$ D5-branes at points on $T^4$.

The D9-branes support an $SO(N)$ gauge field
and the D5-branes can be represented by an $SO(N)$ instanton
on $T^4$.
An $SO(N)$ gauge field can be considered as a connection of
a $U(N)$ bundle $E$ preserving an {\it orthogonal structure}
$I$ of $E$ (i.e. anti-linear isometric involutions $I_x:E_x\to E_x$).
Note that the embedding $SO(N)\hookrightarrow U(N)$ has index $2$,
namely, it maps the generator
of $\pi_3(SO(N))\cong \Z$ to two times the generator of
$\pi_3(U(N))\cong \Z$. Therefore the $2k$ D5-branes,
which are represented by a $2k$-instanton of $U(N)$ group,
correspond to a $k$-instanton of $SO(N)$.

\subsection*{\sl Probing by a Wrapped D5-brane Pair}

We probe this system by a pair of D5-branes wrapped on $T^4$ and
spanning the $x^{\pm}$ directions.
The theory on the probe is
a $(1,0)$ supersymmetric $SU(2)$ gauge theory in $5+1$ dimensions
with $SO(N)$ flavor symmetry
where the eight supersymmetries are broken to half
by the instanton configuration of the flavor group.
The theory has an $SU(2)$ vector multiplet, a singlet hypermultiplet
and a half-hypermultiplet
in the bifundamental representation $({\bf N},{\bf 2}^*)$ 
of $SO(N)\times SU(2)$.
The condition of half-hypermultiplet is important for our purpose
and deserves a paragraph of digression as a reminder.

If a hypermultiplet is in a pseudo-real representation of the
flavor$\times$gauge group, one can impose a half-hypermultiplet condition.
Let $J^{AB}$ be the skew-form defining the psudo-reality
(so that the representation matrix
$g^A_{\,\, B}$ satisfies $J^{AB}\overline{g^B_{\,\, C}}
=g^A_{\,\, B}J^{BC}$ in addition to unitarity),
and let $\epsilon^{\sigma\tau}$ be the invariant tensor of the
$SU(2)_R$ symmetry. Then, the half-hypermultiplet conditions for
the hypermultiplet fields $(Q^{\sigma A},\Psi^A)$ are
\beqa
&&Q^{\sigma A}=J^{AB}\epsilon^{\tau\sigma}\overline{Q^{\tau B}},
\label{halfb}\\
&&\Psi^A=J^{AB}(\Psi^B)^{c_{5\!+\!1}}\label{halff},
\eeqa
where $(\,\cdot\,)^{c_{5\!+\!1}}$ is the charge conjugation in 5+1
dimensions.
\footnote{
In $4n$ dimensional Euclidean or $((4n+1)+1)$ dimensional Minkowski
space, charge conjugation does not flip the chirality.
The square of charge conjugation is $1$ for even $n$ while
it is $-1$ for odd $n$.
The $5+1$ dimensional spinor representation
decomposes to the tensor product
of the $4$ and $1+1$ dimensional spinor representations,
and the $5+1$
dimensional charge conjugation can be represented as the tensor
product of the ones in $4$ and $1+1$ dimensions;
$$(\psi\otimes\lambda)^{c_{5\!+\!1}}
=\psi^{c_4}\otimes\lambda^{c_{1\!+\!1}}.$$

We shall write $(\,\cdot\,)^c$ for $(\,\cdot\,)^{c_4}$.
In the representation where the $d=4$
Gamma matrices are given by
$$
\gamma^{\mu}=\left(
\begin{array}{cc}
0&\sigma^{\mu}\\
\overline{\sigma}^{\mu}&0
\end{array}
\right),~~~\sigma^{\mu}=(i\vec{\sigma},1),~~
\overline{\sigma}^{\mu}=\sigma^{\mu\dag},
$$
the $d=4$ charge conjugation
is given by $\psi^c=C\psi^*$ (on both chirality)
in which $C$ is a charge conjugation matrix given by
$C=i\sigma_2$ and $\psi^*$ is the complex conjugation of
$\psi$.

The $d=1+1$ charge conjugation is simply the complex conjugation
$\lambda^{c_{1\!+\!1}}=\lambda^*$
in the representation where
$\gamma^0=-i\sigma_2$ and $\gamma^9=\sigma_1$ in which
$\gamma^0\gamma^9={\rm diag}(-1,1)$.
In this paper, we often denote $\lambda^*$ by $\overline{\lambda}$.}
In particular, the fermion $\Psi^a$ is a symplectic-Majorana-Weyl
spinor (of negative chirality).
Indeed, the conditions (\ref{halfb}) and (\ref{halff})
are invariant under the (1,0) supersymmetry
\beq
\delta Q^{\sigma A}=\overline{\xi_{\sigma}}\Psi^A,~~~
\delta \Psi^{A}=(\Dsl Q^{\sigma})^A\xi_{\sigma},
\label{susyhyp}
\eeq
generated by a symplectic-Majorana-Weyl spinor
$\xi_{\sigma}=\epsilon_{\sigma\tau}(\xi_{\tau})^{c_{5+1}}$
of positive chirality.
In our case, the flavor$\times$gauge group is
$SO(N)\times SU(2)$ and the representation $({\bf N},{\bf 2}^*)$
is indeed pseudo-real; if we denote the $SU(2)$-gauge indices
by $\rma,\rmb,\ldots$ and the $SO(N)$-flavor indices by $i,j,\ldots$,
the skew-form defining the pseudo-reality is
$\epsilon^{\rma\rmb}\delta^{ij}$.
In the present set-up, the $SO(N)$-flavor bundle is topologically non-trivial
and the hypermultiplet fields are sections $Q^{\sigma \rma}$
and $\Psi^{\rma}$
of the bundle $E$ and $E\otimes S_{\,5\!+\!1}^-$ respectively
where $S_{\,5\!+\!1}^-$ is the $5+1$ dimensional spin bundle
of negative chirality.
Let us define an antilinear map $I^{c_{5\!+\!1}}:
E\otimes S_{\,5\!+\!1}^-\to E\otimes S_{\,5\!+\!1}^-$
as the tensor product of
the map $I$ acting on $E$ and the charge conjugation on
$S_{\,5\!+\!1}^-$. Then, the conditions of half-hypermultiplet
are $Q^{\sigma \rma}
=\epsilon^{\rma\rmb}\epsilon^{\tau\sigma}I(Q^{\tau \rmb})$
and
\beq
\Psi^{\rma}=\epsilon^{\rma\rmb}\, I^{c_{5\!+\!1}}(\Psi^{\rmb}),
\eeq
where $IQ$ is defined by $(IQ)(x)=I_xQ(x)$ and the definition of
$I^{c_{5\!+\!1}}\Psi$ is similar.

At long distances, as in the case without orientifold projection,
we obtain an effective $1+1$ dimensional theory with $(0,4)$
supersymmetry. The $SU(2)$ gauge field on $T^4$ reduces to
scalar fields taking values
in the moduli space of flat $SU(2)$ connections on $T^4$.
A flat $SU(2)$ connection can be represented as a constant gauge field
of the form
\beq
a^{SU(2)}_{\mu}=i\left(
\begin{array}{cc}
a_{\mu}&0\\
0&-a_{\mu}
\end{array}\right).
\label{SU2flat}
\eeq
Since there are $SU(2)$ gauge
equivalence relations
$a_{\mu}\equiv -a_{\mu}
\equiv a_{\mu}+\hat{n}_{\mu}$ ($\hat{n}\in 2\pi \Lambda^*$),
the moduli space is
$(\R^4)^*/(2\pi\Lambda^*\tilde{\times}\Z_2)=\hT^4/\Z_2$.
Away from the $\Z_2$ fixed points of $\hT^4$,
the $SU(2)$ gauge symmetry is
broken to $U(1)$ and we obtain from
the $SU(2)$ vector multiplet
and the singlet hypermultiplet the $1+1$ dimensional
$(0,4)$ multiplets whose bosonic components are
$U(1)$ gauge field $ia_{\pm}$ (embedded in $SU(2)$ as
$a_{\pm}^{SU(2)}={\rm diag}(ia_{\pm},-ia_{\pm})$)
and scalar fields taking values in
$\R^4\times \hT^4$
modulo a $\Z_2$ action which simultaneously flips the sign of
$a_{\pm}$ and the coordinates of $\hT^4$.
At the sixteen $\Z_2$ fixed points, the $SU(2)$ gauge symmetry
is restored.

Massless fields also come from the half-hypermultiplet in the
$({\bf N},{\bf 2}^*)$ representation.
They correspond to the fields on $T^4$ satisfying
the equations as (\ref{zerom})
where now
\beq
D_{\mu}=\partial_{\mu}+A_{\mu}^{SO(N)}-{}^ta^{SU(2)}_{\mu},
\eeq
in which $a^{SU(2)}_{\mu}$ is the flat $SU(2)$
gauge field (\ref{SU2flat}) and $A_{\mu}^{SO(N)}$ is
the $SO(N)$ instanton.
Namely, they come from the zero modes of the Laplace and the Dirac
operators on $T^4$ associated with
$
D_{\mu}^{(\pm a)}=\partial_{\mu}+A_{\mu}^{SO(N)}\mp ia_{\mu}.
$
Generically, there are nothing else than
$2k$ negative-chirality zero modes for each of $\Dsl^{(a)}$
and $\Dsl^{(-a)}$.
As in the previous section,
we expand the field $\Psi^{\rma}$ by the orthonormal basis
$\psi_{I}(\pm a)$ ($I=1,\ldots,2k$)
of the space of zero modes of $\Dsl^{(\pm a)}$ as
\beq
\Psi^1=\sum_{I=1}^{2k}\psi_{I}(a)\otimes \lambda^{1I},~~
\Psi^2=\sum_{I=1}^{2k}\psi_{I}(-a)\otimes \lambda^{2I},
\eeq
where $\lambda^{\rma}=(\lambda^{\rma I})$
are positive chirality spinors in 1+1 dimensions.
The lagrangian for $\Psi^a$ then becomes
\beq
\Bigl(\,\overline{\lambda^1},\,\overline{\lambda^2}\,\Bigr)
\left(\,
\partial_-
+\partial_-a_{\mu}
\left(\begin{array}{cc}
\hat{A}^{\mu}(a)\!&0\\
0&\!\!\!\!-\hat{A}^{\mu}(-a)
\end{array}
\right)
-{}^ta_-^{SU(2)}\,
\right)
\left(\begin{array}{c}
\lambda^1\\
\lambda^2
\end{array}
\right)
\eeq
where $\hat{A}^{\mu}(a)\dd a_{\mu}$
is the $U(2k)$ gauge field on $\hT^4$ (defined as in (\ref{NA}))
defining a connection of the bundle $\hat{E}$ of
$\Dsl^{(a)}$ zero modes.
The bundle $\hat{E}$ has instanton number $N$
and $\hat{A}=\hat{A}^{\mu}(a)\dd a_{\mu}$ has a self-dual
curvature.

The half-hypermultiplet condition constrains the fermions
$\lambda^{\rma}$.
We first note that the operator $I^{c_{5\!+\!1}}$ induces in the
$T^4$ factor a map sending
$\Dsl^{(a)}$ zero modes to $\Dsl^{(-a)}$ zero modes
and vice versa.
Let $I^c:E\otimes S\to E\otimes S$
be the tensor product map of $I$ on $E$
and the charge conjugation on the spinor bundle $S$ on $T^4$.
With respect to a (local) real-orthonormal frame of $E$,
$I^c$ is represented simply as the charge conjugation.
Since
$A_{\mu}^{SO(N)}$ is represented by real anti-symmetric
matrices in such a frame,
the effect of $I^c$ on the Dirac equation
$\gamma^{\mu}(\partial_{\mu}+A_{\mu}^{SO(N)}-ia_{\mu})\psi=0$
is simply to change the sign of $a_{\mu}$.
Namely, if $\psi$ is a zero mode of $\Dsl^{(a)}$, then
$I^c\psi$ (defined as $(I^c\psi)(x)=I^c_x\psi(x)$)
is a zero mode of $\Dsl^{(-a)}$.
Thus, $I^c\psi_{I}(a)$ can be spanned by $\psi_{I}(-a)$'s
\beq
I^c\psi_I(a)=\sum_J\psi_J(-a)J_a^{JI}.
\eeq
Then, the half-hypermultiplet condition
$\Psi^2=I^{c_{5\!+\!1}}\Psi^1$ requires that
\beq
\lambda^{2I}=J_a^{IJ}\overline{\lambda^{1J}}.
\label{halflambda}
\eeq

The anti-linear map $J_a:\hat{E}_a\to \hat{E}_{-a}$ induced by
$\psi\mapsto I^c\psi$
defines a symplectic structure $J$ over the inversion.
Indeed, it squares to $J_{-a}J_a=-1$ due to the property
$\psi^{cc}=-\psi$ of charge conjugation in Euclidean four dimension,
and it is isometric
since $(I^c\psi_1)^{\dag}I^c\psi_2=\psi_2^{\dag}\psi_1$.
A direct computation shows
\beq
J_{a}^{IJ}\overline{\hat{A}^{\mu}(a)_{\bar JK}}
=-\hat{A}^{\mu}(-a)_{\bar IJ}J_a^{JK},
\eeq
which means that the connection $\hat{A}$ preserves $J$.
Thus, we have a symplectic orientibundle $(\hat{E},J)$ over the
orientifold $\hT^4/\Z_2$ with a connection $\hat{A}$.

The theory we have obtained
is exactly the same as the effective theory on a D1-brane pair
probing the system of $k$ pairs of D9-branes in Type IIB orientifold
on $\R^6\times \hT^4/\Z_2$
which support the symplectic orientibundle $(\hat{E},J)$
with the connection $\hat{A}$.
In fact, the condition (\ref{halflambda}) is nothing but
the orientifold projection on 1-9 open string modes.
By definition, this theory must be
identifed as the effective theory of the D1-brane pair which is T-dual to
the original D5-brane pair.
Thus, we conclude
that the T-duality mapping
D9 and D5 branes in Type I string theory on $\R^6\times T^4$
to D9 and D5 branes in Type IIB orientifold on
$\R^6\times \hT^4/\Z_2$ is represented by
the transform $(E,A^{SO(N)},I)\mapsto (\hat{E},\hat{A},J)$
of $SO(N)$ instantons on $T^4$ of instanton number $k$
to $N$-instantons in symplectic orientibundle on $\hT^4/\Z_2$
of rank $2k$.

\subsection*{\sl The Inverse Transform}

The transform
$(E,A^{SO(N)},I)\mapsto (\hat{E},\hat{A},J)$
may be considered as a generalization of
Nahm transform to the case of
orthogonal bundles.
To fully establish this, we should provide the inverse transform.
Thus, we consider D9-branes wrapped on the orthogonal orientifold
$\hT^4/\Z_2$ supporting a rank $2k$ symplectic orientibundle
$(E,A,J)$ of instanton number $N$, and probe this by a wrapped D5-brane.

The theory of lowlying modes on the probe D5-brane can be 
analyzed in the same way as we have done for
wrapped D4-branes in orientifold on
$\R\times (\hT^4\times \R^5)/\Z_2$.
It is a $5+1$ dimensional $(1,0)$ supersymmetric $U(1)$
gauge theory on $\R^2\times \hT^4$
with $U(2k)$ flavor symmetry where the eight
supersymmetries are broken to half by constraints on the fields
(written below) and by the instanton configuration $A^{\mu}\dd
\hx_{\mu}$ of the
flavor group. The theory containes a $U(1)$ vector multiplet,
a free hypermultiplet and a hypermultiplet in the bifundamental
representation $({\bf 2k},-1)$ of $U(2k)\times U(1)$.
The bosonic fields $X^p$ from the free
hypermultiplet and the $U(1)$ gauge field $ia$
(and their superpartners) are subject to the constraints
\beqa
&&X^p(-\hx)=X^p(\hx),\\
&&a^{\mu}(-\hx)=a^{\mu}(\hx),~
a_{\pm}(-\hx)=-a_{\pm}(\hx),
\eeqa
and the $U(1)$ gauge transformations are also constrained as
$g(-\hx)=g(\hx)^{-1}$.
The hypermultiplet in the bifundamental $({\bf 2k},-1)$,
which consists of
sections $Q^{\sigma}$ and $\Psi$
of $E$ and $E\otimes \hat{S}^-_{5\!+\!1}$,
is subject to the ``half-hypermultiplet'' conditions:
$
Q^{\sigma}=\epsilon^{\tau\sigma}JQ^{\tau}
$ and
\beq
\Psi=J^{c_{5\!+\!1}}\Psi,\label{halff2}
\eeq
where $JQ$ and $J^{c_{5\!+\!1}}\Psi$ are defined respectively by
$(JQ)(\hx)=J_{-\hx}Q(-\hx)$ and
$(J^{c_{5\!+\!1}}\Psi)(\hx)
=\Gamma^{1234}J^{c_{5\!+\!1}}_{-\hx}\Psi(-\hx)$
(multiplication by $\Gamma^{1234}$ represents a lift of
the inversion of $\hT^4$ to the spin bundle).
These conditions are invariant under the supersymmetry
(\ref{susyhyp}) generated by the
symplectic-Majorana-Weyl spinor $\xi_{\sigma}$
which is of positive-chirality also in four dimensions
$\Gamma^{1234}\xi_{\sigma}=\xi_{\sigma}$.

The Kaluza-Klein reduction on $\hT^4$
leads to a $1+1$ dimensional $(0,4)$ supersymmetric theory.
The gauge symmetry reduces from $U(1)$ to $\Z_2$ since
$g=\pm 1$ are the only gauge
transformations that are constant along $\hT^4$ satisfying the
constraint.
The vector and the free hypermultiplets reduce to $\Z_2$-singlet
$(0,4)$ multiplets whose bosonic components are scalar fields
taking values in $\R^4\times T^4$
where $T^4$ appears here as the moduli space of (constrained) flat $U(1)$
connections $ix^{\mu}\dd \hx_{\mu}$ on $\hT^4$.
The hypermultiplet in $({\bf 2k},-1)$ leads to $\Z_2$-nonsinglet
positive-chirality
fermions $\lambda^i$ with values in the bundle $\check{E}$ of zero modes
of the Dirac operator associated with the covariant derivative
$D^{\mu}=\partial^{\mu}+A^{\mu}-ix^{\mu}$.
The bundle $\check{E}$ has rank $N$ and instanton number
$2k$.
The fermions $\lambda^i$ are coupled to the dual connection
$\check{A}_{\mu}(x)_{\bar\imath j}$ of $\check{E}$
constructed as before, and are constrained as below.

Let $J^c:E\otimes \hat{S}\to E\otimes\hat{S}$
be the anti-linear map covering the inversion of $\hT^4$
defined as the tensor product
of the map $J:E\to E$ and the charge conjugation followed by
the $\gamma^{1234}$-multiplication
on the spinor bundle $\hat{S}$ of $\hT^4$ (the latter map
is an antilinear lift of the inversion of $\hT^4$
to $\hat{S}$).
In a (local) symplectic frame of $E$, it is
represented as the charge conjugation
followed by the multiplication by $\gamma^{1234}$ and
the matrix $J$ given in (\ref{Jrel}). Then,
the Dirac equation
$\gamma_{\mu}(\partial^{\mu}+A^{\mu}(\hx)-ix^{\mu})
\psi(\hx)=0$ is transformed to
$\gamma_{\mu}(\partial^{\mu}+J\overline{A^{\mu}(\hx)}J^{-1}+ix^{\mu})
\gamma^{1234}J\psi^c(\hx)=0$.
In the symplectic frame, the gauge field $A^{\mu}(\hx)$
satisfies (\ref{SOcond2});
$J\overline{A^{\mu}(\hx)}J^{-1}=-A^{\mu}(-\hx)$.
Thus, we see that $J^c\psi$ (defined as
$(J^c\psi)(\hx)=J^c_{-\hx}\psi(-\hx)$)
is a $\Dsl$ zero mode, if $\psi$ is.
If $\psi_i$ ($i=1,\ldots,N$)
are orthonormal basis of the space of $\Dsl$ zero modes,
$J^c\psi_i$ can be spanned by $\psi_j$'s
\beq
J^c\psi_i=\psi_j\check{I}_x^{ji}.
\eeq
Then, the condition (\ref{halff2}) constrains the fermions $\lambda^i$
(associated with the base $\psi_i$) as
\beq
\lambda^i=\check{I}_x^{ij}\overline{\lambda^j}.
\label{hhaaff}
\eeq

The anti-linear
map $\check{I}_x:\check{E}_x\to\check{E}_x$ induced by
$\psi\mapsto J^c\psi$ defines an orthogonal structure of $\check{E}$.
Indeed, it is an
involution $\check{I}^2=1$ because $\psi^{cc}=-\psi$ and $JJ=-1$,
and also it is an isometry,
$\int J^c\psi_1^{\dag}J^c\psi_2
=\int\psi_2^{\dag}\psi_1$.
It is easy to see that the dual gauge field $\check{A}$ preserves this.
Thus, we have an $SO(N)$ bundle $(\check{E},\check{I})$ with
an $SO(N)$ connection $\check{A}$.

The effective theory we have obtained is exactly the same as the
effective theory of a D1-brane probing the system
of $N$ D9-branes in Type I string theory on $\R^6\times T^4$
which support the $SO(N)$ bundle
$(\check{E},\check{I})$ and the connection $\check{A}$.
Indeed, a Type I D1-brane has a $\Z_2$ gauge symmetry \cite{PW} and also,
(\ref{hhaaff}) is nothing but the orientifold projection
on 1-9 string modes in such a system.
Thus, we can conclude
that the T-duality mapping the D9-D5 system in Type IIB orientifold
on $\R^6\times \hT^4/\Z_2$ to the D9-D5 system in
Type I on $\R^6\times T^4$ is represented by the
transform $(E,A,J)\mapsto (\check{E},\check{A},\check{I})$
of the gauge field configurations.
Since T-duality squares to the identity,
the transforms $(E,A,I)\to (\hat{E},\hat{A},J)$ and
$(E,A,J)\mapsto (\check{E},\check{A},\check{I})$ must be
inverse of each other. This can indeed be shown explicitly
(see Appendix).

\subsection*{\sl Summary}

We have constructed a Nahm transform which is a correspondence of
instantons of orthogonal bundle on $T^4$ and instantons
of symplectic orientibundle on the orthogonal orientifold
$\hT^4/\Z_2$.
The orthogonal bundle of rank $N$ and instanton number $k$
correspond to the symplectic orientibundle of rank $2k$
and instanton number $N$.
By construction, the T-duality between
the D9-D5 system of Type I string theory on $T^4$
and the D9-D5 system of Type IIB on orthogonal orientifold
$\hT^4/\Z_2$ is given by this Nahm transform.

\subsection*{(ii) Symplectic Bundle $\leftrightarrow$
Orthogonal Orientibundle}

We next consider an (unphysical) Type IIB string theory
with $Sp$-type O9-plane on
$\R^6\times T^4$ with $2N$ D9-branes wrapped on $T^4$
and $k$ D5-branes at points on $T^4$.

The D9-branes support an $Sp(N)$ gauge field 
which can be considered as a connection of
a $U(2N)$ bundle $E$ preserving a {\it symplectic structure}
$J$ of $E$ (i.e.
anti-linear isometries $J_x:E_x\to E_x$ such that
$J_x^2=-{\rm id}_x$).
Since the embedding $Sp(N)\hookrightarrow U(2N)$ has index one,
the $k$ D5-branes correspond to a $k$-instanton of $Sp(N)$ on $T^4$.

\subsection*{\sl Probing by a Wrapped D5-brane Pair}

We probe the system by a pair of wrapped D5-branes
as before.
The theory on the probe is a $(1,0)$ supersymmetric $O(2)$
gauge theory in $5+1$ dimensions with $Sp(N)$ flavor symmetry
where the supersymmetries are broken to half
by the instanton configuration of the flavor group.
The theory has an $O(2)$ vector multiplet,
a hypermultiplet in the second rank symmetric tensor representation,
and a half-hypermultiplet
in the bifundamental representation $({\bf 2N},{\bf 2}^*)$ 
of $Sp(N)\times O(2)$.
The conditions of half-hypermultiplet are
$Q^{\sigma \rma}=\epsilon^{\tau\sigma}JQ^{\tau\rma}$
and $\Psi^{\rm a}=J^{c_{5\!+\!1}}\Psi^{\rma}$
where $\rma,\rmb=1,2$ are $O(2)$ gauge indices,
$Q^{\sigma\rma}$ and $\Psi^{\rma}$ are 
the hypermultiplet fields which are sections of the bundle
$E$ and $E\otimes S^-_{5\!+\!1}$ respectively,
and $JQ$ and $J^{c_{5\!+\!1}}\Psi$ are defined as before.

At long distances,
we obtain an effective $1+1$ dimensional theory with $(0,4)$
supersymmetry.
The $O(2)$ gauge field on $T^4$ reduces to
scalar fields taking values
in the moduli space of flat $O(2)$ connections on $T^4$.
A flat $O(2)$ connection can be represented
with respect to a complexified $O(2)$ base
(such that the fermion $\Psi$ is represented as
$\Psi^{\pm}=\Psi^1\mp i \Psi^2$ in the corresponding {\it dual} basis)
as a constant field
\beq
a^{O(2)}_{\mu}=i\left(
\begin{array}{cc}
a_{\mu}&0\\
0&-a_{\mu}
\end{array}\right).
\label{O2flat}
\eeq
Since there are
$O(2)$ gauge equivalence relations
$a_{\mu}\equiv -a_{\mu}
\equiv a_{\mu}+\hat{n}_{\mu}$ ($\hat{n}\in 2\pi \Lambda^*$),
the moduli space is $\hT^4/\Z_2$.
Away from the $\Z_2$ fixed point,
the gauge symmetry is broken to $SO(2)=U(1)$
and the $O(2)$ vector multiplet
plus the symmetric tensor hypermultiplet reduce to
a $U(1)$ gauge field $a_{\pm}^{O(2)}$
and scalar fields taking values in
$\R^4\times \hT^4$
(with their superpartners)
modulo a $\Z_2$ action which flips the sign of
$a_{\pm}^{O(2)}$ and the coordinates of $\hT^4$
(and their superpartners).
At the sixteen fixed points, the $O(2)$ gauge symmetry is restored
and a new branch $\R^4/\Z_2$ develops.

Massless fields also come from the half-hypermultiplet in the
$({\bf 2N},{\bf 2}^*)$ representation.
They correspond to the fields on $T^4$ obeying
the equations as (\ref{zerom})
where now
\beq
D_{\mu}=\partial_{\mu}+A_{\mu}^{Sp(N)}-a^{O(2)}_{\mu},
\eeq
in which $a^{O(2)}_{\mu}$ is the flat
gauge field (\ref{O2flat}) and $A_{\mu}^{Sp(N)}$ is the
$Sp(N)$ instanton.
Namely, they come from the zero modes of the Laplace and the Dirac
operators associated with
$
D_{\mu}^{(\pm a)}=\partial_{\mu}+A_{\mu}^{Sp(N)}\mp ia_{\mu}
$.
Generically
there are nothing else than $k$ Dirac zero modes of
negative-chirality.
We expand
$\Psi^{\pm}=\Psi^1\mp i\Psi^2$
by the orthonormal basis $\psi_I(\pm a)$ ($I=1,\ldots,k$)
of the space of $\Dsl^{(\pm a)}$ zero modes as
$
\Psi^{\pm}=\sum \psi_I(\pm a)\otimes \lambda^{\pm I}
$
where $\lambda^{\pm}=(\lambda^{\pm I})$ are 
$1+1$ dimensional positive-chirality fermions.
Then, the lagrangian for $\Psi^{\pm}$ becomes
\beq
\Bigl(\,\overline{\lambda^{+}},\,\overline{\lambda^{-}}\,\Bigr)
\left(\,
\partial_-
+\partial_-a_{\mu}
\left(\begin{array}{cc}
\hat{A}^{\mu}(a)\!&0\\
0&\!\!\!\!-\hat{A}^{\mu}(-a)
\end{array}
\right)
-a_-^{O(2)}\,
\right)
\left(\begin{array}{c}
\lambda^{+}\\
\lambda^{-}
\end{array}
\right)
\eeq
where $\hat{A}^{\mu}(a)\dd a_{\mu}$
is the $U(k)$ connection of the bundle $\hat{E}$ over $\hT^4$ of
$\Dsl^{(a)}$ zero modes defined as before.
The bundle $\hat{E}$
has instanton number $2N$ and
$\hat{A}=\hat{A}^{\mu}(a)\dd a_{\mu}$ has a self-dual curvature.

The half-hypermultiplet condition
$\Psi^{\pm}=J^{c_{5\!+\!1}}\Psi^{\mp}$ constrains $\lambda^{\pm}$.
Let $J^c:E\otimes S\to E\otimes S$
be the tensor product map of $J:E\to E$
and the charge conjugation on the spinor bundle $S$ on $T^4$.
With respect to a (local) symplectic frame of $E$,
$J^c$ is represented by
the charge conjugation followed by the multiplication
by the $2N\times 2N$ matrix $J$ as (\ref{Jrel}) under which
the Dirac equation
$\gamma^{\mu}(\partial_{\mu}+A_{\mu}^{Sp(N)}-ia_{\mu})\psi=0$
is transformed to
$\gamma^{\mu}(\partial_{\mu}+J\overline{A_{\mu}^{Sp(N)}}J^{-1}+ia_{\mu})
J\psi^c=0$.
In such a frame, the $Sp(N)$ gauge field obeys
$J\overline{A_{\mu}^{Sp(N)}}=A_{\mu}^{Sp(N)}J$.
Thus,
if $\psi$ is a zero mode of $\Dsl^{(a)}$, then
$J^c\psi$ (defined as $(J^c\psi)(x)=J^c_x\psi(x)$)
is a zero mode of $\Dsl^{(-a)}$.
In particular, $J^c\psi_I(a)$ is spanned by
$\psi_I(-a)$'s
\beq
J^c\psi_I(a)=\sum_J\psi_J(-a)I_a^{JI}.
\eeq
Thus, the half-hypermultiplet condition requires
\beq
\lambda^{-I}=I_a^{IJ}\overline{\lambda^{+J}}.
\label{halfspp}
\eeq

The anti-linear map $I_a:\hat{E}_a\to\hat{E}_{-a}$
induced by $\psi\mapsto J^c\psi$
defines an orthogonal structure over the inversion of $\hT^4$.
Indeed, it squares to $I_{-a}I_a=1$
because $\psi^{cc}=-\psi$ and $JJ=-1$, and
is isometric since $(J^c\psi_1)^{\dag}J^c\psi_2=\psi_2^{\dag}\psi_1$.
It is easy to see that
the connection $\hat{A}$ preserves $I$.
Thus, we have an orthogonal orientibundle $(\hat{E},I)$
with a connection $\hat{A}$.

The theory we have obtained
is the same as the effective theory on a D1-brane pair
probing the system of $k$ D9-branes in Type IIB
symplectic-orientifold
on $\R^6\times \hT^4/\Z_2$ which
support the orthogonal orientibundle $(\hat{E},I)$
with the connection $\hat{A}$.
In fact, the condition (\ref{halfspp})
is nothing but the orientifold projection on 1-9 string modes.
Thus, we conclude that the T-duality mapping
the D9-D5 system
in Type IIB symplectic-orientifold on $\R^6\times T^4$
to the D9-D5 system in Type IIB symplectic-orientifold on
$\R^6\times \hT^4/\Z_2$ is represented by
the transform $(E,A^{Sp(N)},J)\mapsto (\hat{E},\hat{A},I)$
of $Sp(N)$ instantons on $T^4$ of instanton number $k$
to $2N$-instantons in orthogonal orientibundle on $\hT^4/\Z_2$
of rank $k$.

\subsection*{\sl The Inverse Transform}

To find the inverse transform,
we consider D9-branes wrapped on the symplectic orientifold
$\hT^4/\Z_2$ supporting an orthogonal orientibundle
$(E,A,I)$ of rank $k$ and instanton number $2N$,
and probe it by a wrapped D5-brane pair.

The theory on the probe
is a $5+1$ dimensional $(1,0)$ supersymmetric $U(2)$
gauge theory on $\R^2\times \hT^4$
with $U(k)$ flavor symmetry where the eight
supersymmetries are broken to half by constraints on the fields
(written below) and by the instanton configuration $A^{\mu}\dd
\hx_{\mu}$ of the
flavor group. The theory containes a $U(2)$ vector multiplet,
an adjoint hypermultiplet and a hypermultiplet in the bifundamental
representation $({\bf k},{\bf 2}^*)$ of $U(k)\times U(2)$.
The bosonic fields $X^p$ from the adjoint
hypermultiplet and the $U(2)$ gauge field $a_{U(2)}$
(and their superpartners) are subject to the constraints
\beqa
&&X^p(-\hx)=\epsilon X^p(\hx)^T\epsilon^{-1},
\label{consSpa}\\
&&a_{U(2)}^{\mu}(-\hx)=\epsilon a_{U(2)}^{\mu}(\hx)^T\epsilon^{-1},~~
a^{U(2)}_{\pm}(-\hx)=-\epsilon a^{U(2)}_{\pm}(\hx)^T\epsilon^{-1},
\label{consSp}
\eeqa
where $\epsilon$ is the $2\times 2$ matrix as $J$ in (\ref{Jrel}).
The gauge transformations are also constrained as
$\epsilon\overline{g(\hx)}=g(-\hx)\epsilon$.
The hypermultiplet in $({\bf k},{\bf 2}^*)$ consisting of
sections $Q^{\sigma\rma}$ and $\Psi^{\rma}$
of $E$ and $E\otimes \hat{S}^-_{5\!+\!1}$ ($\rma=1,2$
are $U(2)$ gauge indices) is subject to the
``half-hypermultiplet'' conditions:
$
Q^{\sigma\rma}
=\epsilon^{\rma\rmb}\epsilon^{\tau\sigma}IQ^{\tau\rmb}
$
and
\beq
\Psi^{\rma}=\epsilon^{\rma\rmb}I^{c_{5\!+\!1}}\Psi^{\rmb},
\label{halffff}
\eeq
where $IQ$ and $I^{c_{5\!+\!1}}\Psi$ are defined by
$(IQ)(\hx)=I_{-\hx}Q(-\hx)$ and $I^{c_{5\!+\!1}}\Psi(\hx)=
\Gamma^{1234}I^{c_{5\!+\!1}}_{-\hx}\Psi(-\hx)$.
These conditions are invariant under the supersymmetry
(\ref{susyhyp}) generated by the
symplectic-Majorana-Weyl spinor $\xi_{\sigma}$
which is positive also in four dimensions
$\Gamma^{1234}\xi_{\sigma}=\xi_{\sigma}$.

The Kaluza-Klein reduction on $\hT^4$
leads to a 1+1 dimensional (0,4) supersymmetric theory.
The $U(2)$ gauge symmetry reduces to $SU(2)$,
since gauge transformations and the gauge field $a_{\pm}^{U(2)}$
that are constant on $\hT^4$ belong to the $SU(2)$ subgroup.
The $\hT^4$ components of the gauge field
reduces to the scalar fields with values in the moduli space
of flat $U(2)$ connection on $\hT^4$ subject
to the constraint (\ref{consSp}).
Such a flat connection can be expressed as
$a_{U(2)}^{\mu}=ix^{\mu}{\bf 1}_2$ where $x^{\mu}$ are parameters,
and the moduli space is $T^4$ because there are gauge equivalence
relations $x^{\mu}\equiv x^{\mu}+n^{\mu}$ ($n\in \Lambda$).
The form of $a_{U(2)}^{\mu}$ shows that these scalar fields are
$SU(2)$ singlets.
The adjoint hypermultiplet subject to (\ref{consSpa})
similarly reduces to $SU(2)$ singlet
free scalar multiplet with values in $\R^4$. 
The bifundamental hypermultiplet
reduces to $SU(2)$-doublet
positive-chirality-fermions $\lambda^{{\rma} i}$
with values in the bundle $\check{E}$ over $T^4$ of zero modes
of the Dirac operator associated with the covariant derivative
$D^{\mu}=\partial^{\mu}+A^{\mu}-ix^{\mu}$.
The bundle $\check{E}$ has rank $2N$ and instanton number $k$.
The fermions $\lambda^{{\rma} i}$
are coupled to the dual connection $\check{A}_{\bar\imath j}$ of
$\check{E}$ constructed as before, 
and are constrained as described below.

Let $I^c:E\otimes \hat{S}\to E\otimes \hat{S}$
be the anti-linear map over the inversion of $\hT^4$
defined as the tensor product
of the map $I:E\to E$ and
the charge conjugation followed by the $\gamma^{1234}$-multiplication
on the spinor bundle $\hat{S}$.
In a (local) real orthogonal frame of $E$, it is simply
represented as the charge conjugation followed by
the $\gamma^{1234}$-multiplication.
This transforms
the Dirac equation
$\gamma_{\mu}(\partial^{\mu}+A^{\mu}(\hx)-ix^{\mu})
\psi(\hx)=0$ to
$\gamma_{\mu}(\partial^{\mu}+\overline{A^{\mu}(\hx)}+ix^{\mu})
\gamma^{1234}\psi^c(\hx)=0$.
In the real orthogonal frame, the gauge field $A^{\mu}(\hx)$
satisfies (\ref{Spcond2});
$\overline{A^{\mu}(\hx)}=-A^{\mu}(-\hx)$.
Thus, we see that $I^c\psi$ (defined as
$(I^c\psi)(\hx)=I^c_{-\hx}\psi(-\hx)$)
is a $\Dsl$ zero mode, if $\psi$ is.
If $\psi_i$ ($i=1,\ldots,2N$) are orthonormal frame
of the space of $\Dsl$ zero modes, $I^c\psi_i$ can be spanned
by $\psi_j$'s
\beq
I^c\psi_i=\psi_j\check{J}_x^{ji}.
\eeq
Then, the condition (\ref{halffff}) constrains the fermions
$\lambda^{{\rma}i}$ (associated with the basis $\psi_i$)
as
\beq
\lambda^{{\rma}i}=\epsilon^{\rma\rmb}\check{J}_x^{ij}
\overline{\lambda^{{\rmb}j}}.
\label{orisppp}
\eeq

The anti-linear map
$\check{J}_x:\check{E}_x\to\check{E}_x$
induced by $\psi\mapsto I^c{\psi}$ 
defines a symplectic structure on $\check{E}$.
Indeed, it squares to $\check{J}^2=-1$
due to $\psi^{cc}=-\psi$, and is isometric since
$I^c\psi_1^{\dag}I^c\psi_2=\psi_2^{\dag}\psi_1$.
It is easy to see that the dual gauge field $\check{A}$ preserves $J$.
Thus, we have an $Sp(N)$ bundle $(\check{E},\check{J})$ over $T^4$
with an $Sp(N)$ connection $\check{A}$.

The effective theory we have obtained is the same as the
effective theory of a D1-brane probing the system
of D9-branes in Type II symplectic orientifold
which support the $Sp(N)$ bundle $(\check{E},\check{A},\check{J})$
on $T^4$.
Indeed, (\ref{orisppp}) is nothing but the orientifold projection
on 1-9 string modes.
Thus, we conclude that the T-duality mapping the
D9-D5 system in orientifold $\hT^4/\Z_2$
to the D9-D5 system in orientifold
on $T^4$ is represented by the transform
$(E,A,I)\mapsto (\check{E},\check{A},\check{J})$
of the gauge field configurations.
Since T-duality squares to the identity, the transforms
$(E,A,J)\mapsto (\hat{E},\hat{A},I)$ and
$(E,A,I)\mapsto (\check{E},\check{A},\check{J})$ must be inverse of
each other. This can indeed be shown explicitly (see Appendix).

\subsection*{\sl Summary}

We have constructed a Nahm transform which is a correspondence of
instantons of symplectic bundle on $T^4$ and instantons
of orthogonal orientibundle on the symplectic orientifold
$\hT^4/\Z_2$.
The symplectic bundle of rank $2N$ and instanton number $k$
correspond to the orthogonal orientibundle
of rank $k$ and instanton number $2N$.
By construction, the T-duality between
the D9-D5 system of Type IIB symplectic orientifold on $T^4$
and the D9-D5 system of Type IIB symplectic orientifold on
$\hT^4/\Z_2$ is given by this Nahm transform.

\subsection*{(iii) Orbibundle}

As the final example,
we consider Type IIB string theory on orbifold
$\R^6\times T^4/\Z_2$ with $2N$ D9-branes wrapped on $T^4/\Z_2$
and $k$ D5-branes at points on $T^4/\Z_2$.
This system is represented by
a rank $2N$ orbibundle $(E,\vph)$ on $T^4/\Z_2$
of instanton number $2k$
with a self-dual connection $A$.

\subsection*{\sl Probing by a Wrapped D5-brane Pair}

We probe the system by a pair of D5-branes wrapped on $T^4/\Z_2$.
The theory of lowlying modes on the probe can be 
analyzed in the same way as before.
It is a $5+1$ dimensional $(1,0)$ supersymmetric $U(2)$
gauge theory on $\R^2\times T^4$
with $U(2N)$ flavor symmetry where the eight
supersymmetries are broken to half by constraints on the fields
(written below) and by the instanton configuration
$A_{\mu}\dd x^{\mu}$ of the
flavor group. The theory containes a $U(2)$ vector multiplet,
an adjoint hypermultiplet and a hypermultiplet in the bifundamental
representation $({\bf 2N},{\bf 2}^*)$ of $U(2N)\times U(2)$.
The bosonic fields $X^p$ from the adjoint
hypermultiplet and the $U(2)$ gauge field $a^{U(2)}$
(and their superpartners) are subject to the constraints
\beqa
&&X^p(-x)=\phi X^p(x) \phi,\\
&&a^{U(2)}_{\mu}(-x)=-\phi a^{U(2)}_{\mu}(x)\phi,~~
a^{U(2)}_{\pm}(-x)=\phi a^{U(2)}_{\pm}(x)\phi,
\label{consP}
\eeqa
where $\phi$ is a $2\times 2$ matrix as $\Phi$ in (\ref{Pdef}).
The gauge transformations are also subject to
the constraint $\phi g(x)\phi=g(-x)$.
The hypermultiplet fields in $({\bf 2N},{\bf 2}^*)$,
the sections $Q^{\sigma\rma}$ and $\Psi^{\rma}$
of $E$ and $E\otimes S^-_{5\!+\!1}$
($\rma\!=\!1,2$ are $U(2)$ gauge indices),
are subject to the ``half-hypermultiplet'' conditions:
$Q^{\sigma\rma}=\phi^{\rma}_{\,\,\rmb}\vph Q^{\sigma\rmb}$
and
\beq
\Psi^{\rma}=\phi^{\rma}_{\,\,\rmb}\vph\Psi^{\rmb},
\eeq
where $\vph Q$ and $\vph\Psi$
are defined respectively by $(\vph Q)(x)=\vph_{-x}Q(-x)$ and
$(\vph \Psi)(x)=\Gamma^{1234}\vph_{-x}\Psi(-x)$.
These conditions are invariant under the supersymmetry
(\ref{susyhyp}) generated by the
symplectic-Majorana-Weyl spinor $\xi_{\sigma}$
which is positive also in four dimensions
$\Gamma^{1234}\xi_{\sigma}=\xi_{\sigma}$.

At long distances, we obtain an effective 1+1 dimensional theory with
$(0,4)$ supersymmetry.
Note that the gauge transformations and the
$\pm$ component of the gauge field
that are constant along $\hT^4$
are in the $U(1)\times U(1)$ subgroup defined by the embedding
\beq
(\e^{is},\e^{it})\mapsto
{1\over 2}\left(
\begin{array}{cc}
\e^{is}+\e^{it}&-\e^{is}+\e^{it}\\
-\e^{is}+\e^{it}&\e^{is}+\e^{it}
\end{array}
\right).
\eeq
Thus, the effective theory has $U(1)\times U(1)$ gauge symmetry
with the gauge field $a_{\pm}^{U(1)\times U(1)}$.
The $T^4$ components of the gauge field reduces to
scalar fields taking values
in the moduli space of flat $U(2)$ connections on $T^4$
subject to the constraint (\ref{consP}).
Such a flat $U(2)$ connection
can be expressed as a constant gauge field of the form
\beq
a^{U(2)}_{\mu}=i\left(
\begin{array}{cc}
a_{\mu}&0\\
0&-a_{\mu}
\end{array}
\right).
\label{U2f}
\eeq
Since there are (constrained) $U(2)$ gauge equivalence relations
$a_{\mu}\equiv -a_{\mu}\equiv a_{\mu}+\hat{n}_{\mu}$
($\hat{n}\in 2\pi \Lambda^*$),
the moduli space is $\hT^4/\Z_2$.
The adjoint hypermultiplet reduces to scalar multiplet
taking values in
the adjoint representation of the $U(1)\times U(1)$ subgroup.
Away from the $\Z_2$ fixed point of $T^4$,
the gauge group
$U(1)\times U(1)$ is broken to its diagonal subgroup $U(1)$,
and we obtain (0,4) supersymmetric $U(1)$ gauge theory
which has singlet scalar fields taking values in $\R^4\times
\hT^4/\Z_2$.
At each of the fixed points,
$U(1)\times U(1)$ is unbroken and
a new branch develops.

In any of these branches, 
massless fields also come from the ``half-hypermultiplet'' in the
$({\bf 2N},{\bf 2}^*)$ representation.
They correspond to the fields on $T^4$ satisfying
the equations as (\ref{zerom})
where now
\beq
D_{\mu}=\partial_{\mu}+A_{\mu}-{}^ta^{U(2)}_{\mu},
\eeq
in which $a^{U(2)}_{\mu}$ is the flat $U(2)$
gauge field (\ref{U2f}).
Namely, they come from the zero modes of the Laplace and the Dirac
operators on $T^4$ associated with
$
D_{\mu}^{(\pm a)}=\partial_{\mu}+A_{\mu}\mp ia_{\mu}.
$
Generically,
there are nothing else than $2k$
negative-chirality zero modes for each of $\Dsl^{(a)}$
and $\Dsl^{(-a)}$.
Let us expand the fermion $\Psi^{\rma}$ by
the orthonormal base $\psi_i(\pm a)$ of the space of
$\Dsl^{(\pm a)}$ zero modes as
\beq
\Psi^1=\sum_i\psi_i(a)\otimes\lambda^{1i},~~~
\Psi^2=\sum_i\psi_i(-a)\otimes\lambda^{2i},
\eeq
where $\lambda^{\rma}=(\lambda^{\rma i})$ are 
$1+1$ dimensional positive-chirality fermions.
The lagrangian for $\Psi^{\rma}$ then becomes
\beq
\Bigl(\,\overline{\lambda^1},\,\overline{\lambda^2}\,\Bigr)
\left(\,
\partial_-
+\partial_-a_{\mu}
\left(\begin{array}{cc}
\hat{A}^{\mu}(a)\!&0\\
0&\!\!\!\!-\hat{A}^{\mu}(-a)
\end{array}
\right)
-{}^ta_-^{U(1)\times U(1)}\,
\right)
\left(\begin{array}{c}
\lambda^1\\
\lambda^2
\end{array}
\right)
\eeq
where $\hat{A}^{\mu}(a)\dd a_{\mu}$
is the $U(2k)$ connection of the bundle $\hat{E}$ over $\hT^2$ of
$\Dsl^{(a)}$ zero modes defined as before.
The bundle $\hat{E}_{\pm}$ have instanton
number $2N$ and $\hat{A}=\hat{A}^{\mu}(a)\dd a_{\mu}$
has a self-dual curvature.

The fermions $\lambda^{\rma}$ are constrained
due to the ``half-hypermultiplet'' condition for $\Psi^{\rma}$.
It is easy to see that, if $\psi$ is a zero mode of $\Dsl^{(a)}$,
then $\vph\psi$ (defined by $(\vph\psi)(x)=\vph_{-x}\psi(-x)$)
is a zero mode of $\Dsl^{(-a)}$.
In particular $\vph\psi_i(a)$ can be spanned by $\psi_j(-a)$:
\beq
\vph\psi_i(a)=\psi_j(-a)\hat{\vph}^{\,\,j}_{a\,\,i}.
\eeq
Then, the ``half-hypermultiplet'' condition requires
\beq
\lambda^{2i}=\hat{\vph}^{\,\,i}_{a\,\,j}\lambda^{1j}.
\label{halforb}
\eeq

The linear map $\hat{\vph}_a:\hat{E}_a\to\hat{E}_{-a}$
induced by
$\psi\mapsto \vph{\psi}$ defines a lift to $\hat{E}$
of the inversion of $\hT^4$.
Indeed, it is involutive and unitary because $\vph$ is.
It is also easy to see that $\hat{A}$ preserves $\hat{\vph}$.
Thus, we have an orbibundle $(\hat{E},\hat{\vph})$ on $\hT^4/\Z_2$
with a connection $\hat{A}$.

The effective theory we have obtained is exactly the same as the
effective theory of a D1-brane probing the system
of $2k$ D9-branes on orbifold $\hT^4/\Z_2$
which support the orbibundle $(\hat{E},\hat{A},\hat{\vph})$.
Indeed, (\ref{halforb}) is nothing but the orientifold projection
on 1-9 string modes.
Thus, we can conclude
that the T-duality mapping the D9-D5 system on orbifold $T^4/\Z_2$
to the D9-D5 system on the dual orbifold $\hT^4/\Z_2$
is represented by the
transform $(E,A,\vph)\mapsto (\hat{E},\hat{A},\hat{\vph})$
of the gauge field configurations.
Since T-duality squares to the identity,
the square of this transform must be identity.
This can indeed be shown explicitly
(see Appendix).

It should be possible to extend
T-duality on orbifold $T^4/\Z_2$ to T-duality on its resolution
--- a smooth K3 surface.
In \cite{HoOz}, the transformation of D-branes under
T-duality on K3 surface was proposed using Mukai's Fourier
transform. It is interesting to derive it using our argument and
to see the relation to the construction of the present section.
See also \cite{RW} for a related discussion which focuses on
the B-field.

\section{Reduction to Topology}

Recently,
it was argued by Witten that D-brane charges take values
in K-theory groups of the space-time \cite{WittenK}.
Since T-duality is an equivalence of string theories which sends
D-branes to D-branes, (if the identification of D-brane charge as
K-theory element is valid for any size of the space-time)
it should induce an isomorphism of relevent K-theory groups.
One is then interested in what this isomorphism is
in general.
In this section, we determine this for T-duality on four-torus
by reducing to topology the Nahm transforms obtained
in the previous sections.
This serves as a warm-up for the next section where we will determine
the isomorphisms for tori of other dimensions.

\subsection{D-branes, K-Theory, and Index Theory}

The basic assumption behind the identification of K-groups
as D-brane charges \cite{WittenK}
is that a D$p$-brane and an anti-D$p$-branes can annihilate
by condensation of the tachyon field.\footnote{The tachyon
is created by stretched strings 
and its condensation breaks the off-diagonal $U(1)$ subgroup
of $U(1)\times U(1)$. There has been a puzzle about the fate of
diagonal $U(1)$ which appears to remain unbroken \cite{Sred,WittenK}.
However, it can disappear by confinement, namely,
by condensation of the ``magnetic tachyon'' which is created by
stretched (anti-)D$(p-2)$-branes and is charged under
the $(p-2)$-form potential
dual to the diagonal $U(1)$.
See \cite{LeeYi} for some of the details.
In the present context, these are simply invisible
sectors and will not be mentioned.}
D-branes located at a submanifold $W$ of the space-time
support a complex vector bundle $E$ on $W$ with a connection.
Likewise anti-D-branes at $W$
support another vector bundle $F$ with a connection.
The tachyon field is a complex linear map $T:E\to F$
(and its conjugate $T^{\dag}:F\to E$).
If $T$ is everywhere at the minimum of the tachyon potential
and hence is an isomorphism,
then the system is considered to be equivalent to the vacuum;
i.e. if $E$ and $F$ are isomorphic, the
branes and anti-branes will annihilate.
The set of pairs $(E,F)$ modulo an equivalence relation
$(E,F)\equiv (E^{\prime},F^{\prime})$
when there are $H$ and $H^{\prime}$ such that
$(E\oplus H,F\oplus H)\cong
(E^{\prime}\oplus H^{\prime},F^{\prime}\oplus H^{\prime})$
forms the K-theory group $\K(W)$.

We have actually experienced in section 2 a phenomenon analogous to
brane-anti-brane annihilation; it is the decoupling
of brane-anti-brane pairs in the infra-red limit of the probe theory. 
We considered a vector bundle $E$ on $T^4$ supported
by D9-branes and probed the system with a D5-brane wrapped on $T^4$.
The effective theory on the D5-brane, which is a 1+1
dimensional theory with a tower of Kaluza-Klein modes,
can be identified with the effective theory
of a D1-brane probing the T-dualized system.
The infinite Kaluza-Klein modes from the 5-9 hypermultiplet
are interpreted as the strings stretched between the D1-brane
and infinitely-many D9 or anti-D9-branes of the T-dualized system.
The fermion masses come from the Dirac operators on $T^4$
and are interpreted as the tachyon vevs of infinite D9 anti-D9 pairs.
Massive modes are irrelevant at long distances and can be
simply ignored. We can interpret this as the pair annihilation of
the D9 and anti-D9-branes via tachyon condensation.

Now, as in \cite{WittenK}, we can associate to this
system of D9 and anti-D9-branes an element of
an appropriate K-theory group which can be considered as
the D-brane charge of the T-dualized system.
Let ${\cal E}^+$
and ${\cal E}^-$ be the spaces of
sections of negative and positive-chirality spinor
bundles on $T^4$ coupled to $E$, as in section 2.
The Dirac operator
$\Dsl$ defines a complex linear map
$\D:{\cal E}^+\to{\cal E}^-$ and its conjugate
$\D^{\dag}:{\cal E}^-\to{\cal E}^+$.
The spaces ${\cal E}^+$ and ${\cal E}^-$ together with
the operator $\D$ are parametrized by $a_{\mu}\in \hT^4$
and are considered as bundles
over $\hT^4$ and a map between them.
We have interpreted them
as the bundles supported by the D9 and the
anti-D9-branes and the tachyon field of the T-dualized system.
Thus, the desired element of a K-theory group is
``$({\cal E}^+,{\cal E}^-)$'' which belongs to $K(\hT^4)$.

One may wonder whether
the fact that the bundles ${\cal E}^+$ and ${\cal E}^-$
are infinite-dimensional causes some trouble.
However, since the massive modes are irrelevant
(or since brane-anti-brane pair can
annihilate via tachyon condensation),
we can throw away the higher level modes
and reduce the problem to finite dimension.
Let us decompose ${\cal E}^+$ and ${\cal E}^-$ as
${\cal E}^+={\cal E}^+_0\oplus {\cal E}^+_1$ and
${\cal E}^-={\cal E}^-_0\oplus {\cal E}^-_1$ in such a way that
$\D$ sends ${\cal E}^+_i$ to ${\cal E}^-_i$ ($i=0,1$),
${\cal E}^+_0$ and ${\cal E}^-_0$ are finite-dimensional,
and $\D:{\cal E}^+_1\to{\cal E}^-_1$ is an isomorphism.
Locally such a decomposition always exists as one can see, say, by
taking the spectral decomposition and defining
${\cal E}^+_1$ and ${\cal E}^-_1$ as the modes whose Dirac eigenvalues
do not vanish. If we could take such a
decomposition globally on $\hT^4$, we would be able to regularize
``$({\cal E}^+,{\cal E}^-)$'' by $({\cal E}^+_0,{\cal E}^-_0)$
defining an element of $K(\hT^4)$.
Such a global decomposition does not always exist,
but we can glue the local decompositions
to obtain an element of $K(\hT^4)$ which restricts locally to
$({\cal E}^+_0,{\cal E}^-_0)$.
\footnote{Here is the construction:
Collecting the spaces ${\cal E}^-_0$ defined locally,
using cut-off functions
we obtain a finite-dimensional vector bundle $V$ over $\hT^4$
with a map $f:V\to {\cal E}^-$ such that
$\D\oplus f:{\cal E}^+\oplus V\to{\cal E}^-$ is surjective
everywhere.
The kernel of $\D\oplus f$ has a constant rank
and defines a vector bundle over $\hT^4$.
Then, we define ``$({\cal E}^+,{\cal E}^-)$'' by
$({\rm Ker}(\D\oplus f),V)$.}
This is actually exactly what the family index theory
defines \cite{ASIV,ASV,ASAd}
as the index of $\D:{\cal E}^+\to {\cal E}^-$.

The operator $\D:{\cal E}^+\to {\cal E}^-$
is defined as the Dirac operator $D^+:\Gamma(S^+)\to \Gamma(S^-)$
coupled to the family of connections
$A_{\mu}-ia_{\mu}$ over $T^4$ parametrized by
$a\in \hT^4$.
Its index actually depends only on
the topology of the bundle ${\cal V}$ on $T^4\times \hT^4$
which carries this family of connections
(i.e. the bundle which has a connection which restricts on
$T^4\times \{a\}$ to the connection $A_{\mu}-ia_{\mu}$).
The index can therefore be denoted as $\ind(D^+\!,{\cal V})$.
Thus, we need to find out what ${\cal V}$ is.

\subsection*{\sl The Poincar\'e Bundle}

The $U(1)$ connection $\partial_{\mu}-ia_{\mu}$ on $T^4$ is
equivalent to $\partial_{\mu}-i(a_{\mu}+\hat{n}_{\mu})$ for
$\hat{n}_{\mu}\in2\pi\Lambda^*$ under the gauge transformation
$\e^{i\hat{n}x}$.
Also, we recall from section 2 that
the $U(1)$ gauge symmetry of the probe D5-brane implies that
the section $\psi(x)$ of $S^{\pm}\otimes E$
at $a\in (\R^4)^*$ should be identified
as the section $\e^{i\hat{n}x}\psi(x)$
at $a+\hat{n}\in (\R^4)^*$.
These motivate us to define a complex line bundle
${\cal P}$ over $T^4\times \hT^4$
as the quotient of the trivial bundle
$T^4\times (\R^4)^*\times \C$ by the action of
the lattice $2\pi\Lambda^*$ given by
\beq
\hat{n}:(x,a,c)\longmapsto (x,a+\hat{n},\e^{i\hat{n}x}c).
\label{act}
\eeq
We call this the {\it Poincar\'e bundle}.
The gauge potential $-ia_{\mu}\dd x^{\mu}$ on
$T^4\times (\R^4)^*$ is invariant under the transformation
(\ref{act}) and defines
a connection $\omega$ of ${\cal P}$ that restricts on $T^4\times\{a\}$
to the flat connection $\partial_{\mu}-ia_{\mu}$ of the trivial bundle.
When restricted to $\{x\}\times \hT^4$, the connection $\omega$
yields a flat connection on $\hT^4$ which is equivalent to
$\partial/\partial a_{\mu}+ix^{\mu}$.
However, $\omega$ is not flat over $T^4\times\hT^4$
and has a curvature $\dd \omega$ with a first Chern class
\beq
c_1({\cal P})=\sum_{\mu=1}^4{\dd a_{\mu}\wedge \dd x^{\mu}
\over 2\pi}=\sum_{i=1}^4\widehat{\eta}^i\wedge\eta_i,
\label{c1P}
\eeq
where $\eta_i$ and $\widehat{\eta}^i$ are basis of
$H^1(T^4,\Z)$ and $H^1(\hT^4,\Z)$ respectively
(more precisely, their image under the embedding of the integral to
the real cohomology of the torus).

\subsection*{\sl The Isomorphism $\K(T^4)\,\cong\,\K(\hT^4)$}

The tensor product $E\otimes {\cal P}$ has a connection 
$A\otimes 1+1\otimes \omega$
that restricts on $T^4\times \{a\}$ to the connection
$A_{\mu}-ia_{\mu}$.
Also, ${\cal E}^+$ and ${\cal E}^-$ are vector bundles whose
fibres at $a\in \hT^4$ are given by
${\cal E}^+_a=\Gamma(T^4,S^+\otimes E\otimes {\cal P}|_a)$
and ${\cal E}^-_a=\Gamma(T^4,S^-\otimes E\otimes {\cal P}|_a)$.
Therefore, $E\otimes {\cal P}$
is the desired bundle over $T^4\times \hT^4$
that carries the family of connections
defining the operator $\D:{\cal E}^+\to {\cal E}^-$.

Thus, we find that the T-duality transformation of the D-brane charge
is given by $E\mapsto E\otimes {\cal P}\mapsto
\ind(D^+\!,E\otimes {\cal P})$, which can be considered as the
image of $(E,0)\in K(T^4)$ under the composition of the maps
of K-theory groups:
\beq
\K(T^4)\,\stackrel{\otimes{\cal P}}{-\!\!\!\longrightarrow}\,
\K(T^4\times \hT^4)\,\stackrel{\ind D^+}{-\!\!\!\longrightarrow}\,
\K(\hT^4).
\label{topNahm}
\eeq
By abuse of language, we shall denote $\ind(D^+,E\otimes{\cal P})$
by $\widehat{E}$.
Since the Nahm transform squares to the identity transform of instantons,
the above map, which is the topological reduction of the Nahm
transform, must also square to the identity of $\K(T^4)$.
Namely, the inverse of
(\ref{topNahm}) is given by
\beq
\K(T^4)\,\stackrel{\ind\widehat{D}^+}{\longleftarrow\!\!\!-}\,
\K(T^4\times \hT^4)\,
\stackrel{\otimes\widehat{\cal P}}{\longleftarrow\!\!\!-}\,
\K(\hT^4),
\label{topNahmback}
\eeq
where $\widehat{D}^+$ is the Dirac operator of $\hT^4$ and
$\widehat{\cal P}$
is the dual of ${\cal P}$ with connection
$-\omega$.

To compute the index, we use the formula
\beq
\ch(\ind(D^+\!,{\cal V}))=
\int\limits_{X}\ch({\cal V})\widehat{A}(X)
\label{evenD}
\eeq
which holds for the Dirac operator $D^+:\Gamma(S^+)\to \Gamma(S^-)$
on any even-dimensional spin manifold $X$
 coupled to any family
of connections carried by
a bundle ${\cal V}$ over $X\times Y$ where $Y$ is a compact
parameter space.
Here, $\widehat{A}(X)$ is the A-roof genus of $X$ which belongs to
$H^*(X,\Q)$
and $\ch$ is the Chern character, which is a group homomorphism
\beq
\ch: \K(X)\to H^{\rm even}(X,\Q)
\label{ch}
\eeq
defined for a general topological space $X$.
Note that $\ch_0=\rank$, $\ch_1=c_1$, and $\ch_2=c_1^2/2-c_2$ etc.

In the present case, we have (see the next section)
\beq
\K(T^4)\cong\Z^8
\eeq
and the Chern character maps this isomorphically to the subgroup
$H^{\rm even}(T^4,\Z)\cong\Z^8$ of $H^{\rm even}(T^4,\Q)\cong\Q^8$
where $H^0(T^4,\Z)\cong\Z$, $H^2(T^4,\Z)\cong\Z^6$ (generated by
$\eta^i\eta^j$), $H^4(T^4,\Z)\cong\Z$ (generated by the volume form). 
In particular, no information is lost by looking only at the Chern
character.
Since we have ${\cal V}=E\otimes {\cal P}$ and $\widehat{A}(T^4)=1$
the character of the index is
$\ch(\widehat{E})=\int_{T^4}\ch(E)\ch({\cal P})$.
Since ${\cal P}$ is a line bundle we have
$\ch({\cal P})=\e^{c_1({\cal P})}$.
Then, (\ref{c1P}) shows that the Chern numbers of $E$ are related
to that of the index $\widehat{E}$ as
\beqa
&&\rank(\widehat{E})=\ch_2(E),\\
&&c_1(\widehat{E})=-\sigma(c_1(E)),\\
&&\ch_2(\widehat{E})=\rank(E),
\eeqa
where $\sigma$ is a map of $H^2(T^4,\Z)$ to $H^2(\hT^4,\Z)$
sending $\eta^i\eta^j$ to
${1\over 2}\epsilon_{ijkl}\widehat{\eta}^k\widehat{\eta}^l$.
In particular, the map $\K(T^4)\to\K(\hT^4)$ given by (\ref{topNahm})
is an isomorphism.

\subsection{Orthogonal/Symplectic (Orienti)bundles and $\Z_2$
Orbibundles}

We next find the map of D-brane charges under the T-duality
on $T^4$ in the presence of orientifold/orbifold projection,
by considering
the topological reduction of the Nahm transforms
constructed in section 3.

\subsection*{\sl Structures of the Poincar\'e Bundle}

The bundle ${\cal P}$ and the connection $\omega$
has two basic properties which
are useful for the present discussion (and which we have actually
implicitly used in the construction of the Nahm transform).
One is the structure of an orthogonal orientibundle.
We note that the anti-linear involution
\beq
I: (x,a,c)\mapsto (x,-a,\overline{c})
\eeq
of the trivial bundle over $T^4\times (\R^4)^*$
commutes with the action (\ref{act}) of $2\pi\Lambda^*$, and hence
defines and orthogonal structure on ${\cal P}$
over $(x,a)\mapsto (x,-a)$. We also note that the connection $\omega$
coming from the gauge potential $-ia_{\mu}\dd x^{\mu}$
is invariant under this $\Z_2$ action.
Thus, $({\cal P},\omega,I)$ becomes an orthogonal orientibundle
with a connection over $T^4\times\hT^4$ with respect to the
inversion of the $\hT^4$ factor.
The other is the structure of an orbibundle.
The involution
\beq
\varphi: (x,a,c)\mapsto(-x,-a,c)
\eeq
obviously commutes with the $2\pi\Lambda^*$ action
and preserves the connection $\omega$.
Thus, $({\cal P},\omega,\varphi)$ is an orbibundle with a connection
over $T^4\times \hT^4$ with respect to the total inversion.

\subsection*{\sl Relevant K-theory Groups}

As we have seen in section 3.3,
in theories with orientifold/orbifold projection 
the Chan-Paton bundles on D9-branes have various
extra structures.
Accordingly, the relevant K-theory groups vary.

The K-theory group for orthogonal bundles over a space
$X$ is the KO-theory group of $X$ denoted by $\KO(X)$. 
That for symplectic bundles is likewise denoted by
$\KSp(X)$.
Now let $Y$ be a space with an involution $\sigma$.
Then, the K-theory group for orthogonal orientibundles
over $Y$ with respect to $\sigma$ is what is known as
KR-theory group and is denoted as $\KR(Y)$.
We can also define K-theory group for
symplectic orientibundles over $Y$ with respect to $\sigma$.
We shall denote it by $\KPR(Y)$.\footnote{The same group
appears in \cite{Gukov} and is named as $KH(Y)$.}
Finally, the K-theory for $\Z_2$ orbibundles over $Y$
with respect to $\sigma$ is given by the
$\Z_2$-equivariant K-theory and the group is denoted by $\KZt(Y)$.

Note that orthogonal (orienti)bundles and $\Z_2$-orbibundles
are closed
under tensor product. Therefore $\KO(X)$, $\KR(Y)$ and $\KZt(Y)$
become rings.
However, symplectic (orienti)bundles are not closed
and thus $\KSp(X)$ and $\KPR(Y)$ has no ring structure.
Instead, the tensor product of two symplectic (orienti)bundles is
an orthogonal (orienti)bundle, and the tensor product of
an orthogonal and a symplectic (orienti)bundles is a symplectic
(orienti)bundle.
Thus, the sums $\KO(X)\oplus\KSp(X)$ and $\KR(Y)\oplus\KPR(Y)$
become rings.

\subsection*{(i) $\KO(T^4)\,\cong\, \KPR(\hT^4)$}

We first consider the $SO$-type orientifold
and find the transformation of D-brane charges under T-duality on
four-torus.
As we have seen in section 3, D9-branes in Type I string theory
support orthogonal bundles while those in Type IIB
orientifold on $\hT^4/\Z_2$ support symplectic orientibundles
over $\hT^4$ with respect to the inversion.
Thus, we must find a map from $\KO(T^4)$ to $\KPR(\hT^4)$
and back.
As in the previous subsection, we can find this
by taking the topological reduction of the Nahm transform
between orthogonal bundles and symplectic orientibundles
obtained in section 3.
We recall that the transform
is based on the Nahm transform for the
underlying unitary bundles.
Thus, we basically follow the maps (\ref{topNahm})
and (\ref{topNahmback}).

Since the Poincar\'e bundle ${\cal P}$ has an orthogonal structure
over the inversion of $\hT^4$, the tensor product of an orthogonal
bundle over $T^4$ (pulled back to $T^4\times \hT^4$)
and ${\cal P}$ defines an orthogonal orientibundle over
$T^4\times \hT^4$ with respect to the inversion of the
$\hT^4$ factor.
Thus, $\otimes {\cal P}$ defines a map from $\KO(T^4)$
to $\KR(T^4\times \hT^4)$.
On the other hand, the spinor bundles $S^{\pm}$
on $T^4$ (or on any four-dimensional spin manifold)
has a symplectic structure coming from the charge conjugation.
Thus, tensoring an orthogonal orientibundle with $S^{\pm}$
(pulled back to $T^4\times \hT^4$) makes a symplectic orientibundle.
Then, the index of the Dirac operator on $T^4$ becomes
a symplectic orientibundle over $\hT^4$.
Thus, $\ind D^+$ defines a map from
$\KR(T^4\times\hT^4)$ to $\KPR(\hT^4)$.
By composition, we obtain
a map
\beq
\KO(T^4)\,\stackrel{\otimes{\cal P}}{-\!\!\!\longrightarrow}\,
\KR(T^4\times \hT^4)\,\stackrel{\ind D^+}{-\!\!\!\longrightarrow}\,
\KPR(\hT^4).
\label{topONahm}
\eeq
In section 3, we have seen that
the symplectic structure over the inversion of
$\hT^4$ obtained this way is exactly the same as
the one for the T-dualized system.
Thus, (\ref{topONahm}) is the desired map for the T-duality
transformation of the D-branes charges.

The inverse map can be obtained in a similar way.
We only have to note that
the dual Poincar\'e bundle
$\widehat{\cal P}$ also has an orthogonal structure
over the inversion of the $\hT^4$ factor and that
the spinor bundles $\widehat{S}^{\pm}$ on $\hT^4$
has a symplectic structure over the inversion coming from the charge
conjugation followed by the multiplication by $\Gamma^{1234}$.
The result is
\beq
\KO(T^4)\,\stackrel{\ind\widehat{D}^+}{\longleftarrow\!\!\!-}\,
\KPR(T^4\times \hT^4)\,
\stackrel{\otimes\widehat{\cal P}}{\longleftarrow\!\!\!-}\,
\KPR(\hT^4).
\label{topONahmback}
\eeq

The two maps (\ref{topONahm}) and (\ref{topONahmback}) must be the
inverse of each other,
since the Nahm transforms are.
In particular, these must be isomorphisms of groups.
Indeed these groups are isomorphic (see the next section)
\beq
\KO(T^4)=
\Z\oplus\Z_2^{\oplus 4}\oplus\Z_2^{\oplus 6}\oplus\Z
=\KPR(\hT^4).
\eeq

\subsection*{(ii) $\KSp(T^4)\,\cong\, \KR(\hT^4)$}

We next consider T-duality on four-torus of
the $Sp$-type orientifold.
As before, we
obtain the map of D-brane charges,
from $\KSp(T^4)$ to $\KR(\hT^4)$ and back,
by topological reduction of the Nahm transform of section 3.
We use the orthogonal structure of ${\cal P}$ ($\widehat{\cal P}$)
over the inversion of the $\hT^4$ factor
and the symplectic structure of the spinor bundles $S^{\pm}$
($\widehat{S}^{\pm}$).
The result is
\beq
\KSp(T^4)\,\stackrel{\otimes{\cal P}}{-\!\!\!\longrightarrow}\,
\KPR(T^4\times \hT^4)\,\stackrel{\ind D^+}{-\!\!\!\longrightarrow}\,
\KR(\hT^4),
\eeq
and the inverse is
\beq
\KSp(T^4)\,\stackrel{\ind \widehat{D}^+}{\longleftarrow\!\!\!-}\,
\KR(T^4\times \hT^4)\,
\stackrel{\otimes\widehat{\cal P}}{\longleftarrow\!\!\!-}\,
\KR(\hT^4).
\eeq
The two groups are indeed isomorphic
\beq
\KSp(T^4)=\Z\oplus\Z=\KR(\hT^4).
\eeq

\subsection*{(iii) $\KZt(T^4)\,\cong\, \KZt(\hT^4)$}

Finally we obtain the map from $\KZt(T^4)$ to $\KZt(\hT^4)$
by reducing the Nahm transform.
What we use here is the fact that the Poincar\'e bundle
${\cal P}$ has a complex linear lift of the inversion of
$T^4\times \hT^4$ and that the spinor bundles $S^{\pm}$
also have a linear lift of the inversion
given by the multiplication by $\Gamma^{1234}$.
Then, the map is easily obtained. It is
\beq
\KZt(T^4)\,\stackrel{\otimes{\cal P}}{-\!\!\!\longrightarrow}\,
\KZt(T^4\times \hT^4)\,\stackrel{\ind D^+}{-\!\!\!\longrightarrow}\,
\KZt(\hT^4).
\eeq
The inverse map is similar.

\section{Generalizations}

In the final section, we identify the isomorphism of
appropriate K-theory groups that realizes
the map of D-brane charges
under T-duality on torus of arbitrary dimensions.

\subsection{Relevant K-theory Groups}

As argued in \cite{WittenK} and as we have experienced,
D-branes that can be represented as
bound states of D$p$-branes and anti-D$p$-branes wrapped on
a $(p+1)$-dimensional submanifold $W$ are classified by
an appropriate K-theory group of $W$.
In string thoery, it seems common that {\it any} D-brane
can be represented as a bound state of the highest dimensional
D-branes (and anti-D-branes) filling the ten-dimensional space-time $X$.
Therefore, the {\it entire} D-brane charges of a string theory on $X$
can be classified by an appropriate K-theory group
of the full space-time $X$.

The relevant K-theory group for Type IIB string theory on $X$
is $\K(X)$ while it is $\KO(X)$
for Type I string on $X$ \cite{WittenK}.
It was also proposed in \cite{WittenK}
that D-brane charge in Type IIA string theory
takes values in $\K^{-1}(X)$ --- the subgroup of
$\K(S^1\times X)$ consisting of elements that are
trivial on $z_0\times X$ where $z_0$ is a point in $S^1$.
This proposal was supported by Ho\v rava in \cite{Horava}.
He first argued that 
any Type IIA (BPS) D-brane can possibly be represented as
a certain bound state of the unstable non-supersymmetric
D9-branes \cite{Sen12,Horava}
and then identified the Chan-Paton bundle $E$
with the tachyon field $T:E\to E$ of D9-branes
as the pair $(E,\e^{i T})$ which determines an element of $\K^{-1}(X)$
via another version of the definition of $\K^{-1}(X)$.
See \cite{Gukov} for another discussion.

However, the precise K-theory group
has not yet been identified for Type II orientifold
except for the cases where one knows
the Chan-Paton bundles on ninebranes; for example
it is $\KR(X)$ ($\KR^{-1}(X)$)
when the Chan-Paton bundles on D9 and anti-D9-branes (IIA ninebranes)
are orthogonal orientibundles \cite{WittenK}(\cite{Horava}).\footnote{
It was also proposed in \cite{WittenK} that it is
$\KR_{\pm}(X)$ for a different kind of Chan-Paton bundles.
K-theory for orientifolds was
also studied in \cite{Gukov} with an extension
to symplectic Chan-Paton bundles.}
Here, we consider the $\Z_2$ orientifold group generated
by an involution $\tau$ of the space-time $X$ times the
worldsheet orientation reversal.\footnote{The generator is
$(-1)^{c_pF_L}\tau\Omega$ with $c_p={(9-p)(8-p)\over 2}$
in the standard notation,
where we assume that there are O$p$-planes
for a single $p$ --- odd for IIB, even for IIA
($(-1)^{c_p F_L}$ is for the group to be $\Z_2$).
As before, an O$p$-plane $W$
is a $(p+1)$-dimensional submanifold of $X$
consisting of fixed points of the involution $\tau$
which acts near $W$ as the sign flip of $(9-p)$ normal coordinates.
It is of $SO$-type or $Sp$-type depending on the sign of the
surrounding $\R{\rm P}^2$ diagram of fundamental string.}
D$p$-branes on top of an O$p$-plane $W$
support orthogonal Chan-Paton bundle over $W$
if $W$ is of $SO$-type
while they support symplectic bundle over $W$
if $W$ is of $Sp$-type.
Away from orientifold planes, physics in the quotient space
$X/\Z_2$ is locally the same as
the underlying Type II string theory
and in particular D-branes locally support unitary bundles.
Thus, the relevant K-theory group of $X$ must satisfy the two
conditions:

\noindent
(i)\, The group corresponding to D-branes localized on an orientifold
plane $W$ must be $\KO(W)$ or $\KSp(W)$
 if $W$ is of $SO$-type or of $Sp$-type respectively.\\
(ii)\, For a region $R$ in $X/\Z_2$ covered by
two separate copies of $R$ in $X$,
the group corresponding to D-branes in $R$
must be $\K(R)$ for IIB
and $\K^{-1}(R)$ for IIA orientifold.

We shall propose K-groups which satisfy these conditions.
In order to present it, we need to introduce some notions and
facts in K-theory. See \cite{Atiyah,AtiReal,Karoubi} for details.

\subsection*{\sl The Groups $\KR^{-n}(Y)$ and $\KPR^{-n}(Y)$}

Let $Y$ be a space with an involution $\sigma$.
Let $D^{p,q}$ be the unit disc
in $\R^{p+q}$
on which we let the involution act trivially on the
first $p$ coordinates but by sign-flip on the last $q$ coordinates.
We consider
orthogonal orientibundles over $D^{p,q}\times Y$
with respect to the involution
 acting on $Y$ by $\sigma$
and on $D^{p,q}$ as dictated above.
We define the group
$\KR^{-p,-q}(Y)=\KR(D^{p,q}\times Y,\partial D^{p,q}\times Y)$ as
the KR-group for bundles over
$D^{p,q}\times Y$ that vanish on $\partial D^{p,q}\times Y$.
Here ``bundle over $A$ that vanishes on $B\subset A$'' stands for
a pair $(E,F)$ where $E$ and $F$ are
bundles over $A$ that are isomorphic on $B$, $E|_B\cong F|_B$.
Replacing ``orthogonal'' by ``symplectic'' we obtain
the group $\KPR^{-p,-q}(Y)$.
We denote
$\KR^{-n}(Y)=\KR^{-n,0}(Y)$ and $\KPR^{-n}(Y)=\KPR^{-n,0}(Y)$.
The following relations between these groups are known as Bott
periodicity
\beqa
&&\KR^{-p,-q}=\KR^{-p-1,-q-1},\label{Bott1}\\
&&\KR^{-n}=\KR^{-n-8},\label{Bott2}\\
&&\KPR^{-n}=\KR^{-n\pm 4}.\label{Bott3}
\eeqa
Using these, we define $\KR^n$ etc for $n>0$.
We see that any of these groups is equal to
$\KR^{-n}$ for some $n$.
We also define $\KR^{-n}(Y,Z)$
for a $\Z_2$-invariant subspace $Z\subset Y$
as the $\KR^{-n}$-group for
bundles over $Y$ that vanish on $Z$.
\footnote{More precisely, $\KR^{-n}(Y,Z)$ is the KR-group
for bundles over $S^n\wedge (Y/Z)$
that vanish on the base point,
where $A\wedge B$ is the smash product defined by
$(A\times B)/((A\times b_0)\cup(a_0\times B))$
(see \cite{Atiyah}).}

When the involution $\sigma$ acts trivially on
$Y$, we have $\KR(Y)=\KO(Y)$ and 
$\KPR(Y)=\KSp(Y)$, and we define $\KO^{-n}(Y):=\KR^{-n}(Y)$
and $\KSp^{-n}(Y):=\KPR^{-n}(Y)$.
When $Y$ is a disjoint union of two copies of a space $R$
and the involution $\sigma$ is the
exchange of the copies,
we have $\KR^{-n}(Y)=\K^{-n}(R)$ where $\K^{-n}$
is defined in a similar way and obeys Bott periodicity
$\K^{-n}=\K^{-n-2}$
so that $\K^{-n}=\K$ for even $n$ and $\K^{-n}=\K^{-1}$ for odd $n$.

\subsection*{\sl The proposal}

We now propose the relevant K-theory group:
{\it D-brane charges of
Type II orientifold on $X/\Z_2$ are classified by}
\beqa
\KR^{-(9-p)}(X)~\,
\mbox{if there are O$p$-planes of $SO$-type only},\\ 
\KPR^{-(9-p)}(X)~\,
\mbox{if there are O$p$-planes of $Sp$-type only.}
\eeqa

Let us show that this proposal satisfy the conditions (i) and
(ii). We only consider $SO$-type orientifold (since $Sp$-type
orientifold is similar) and we put $n=9-p$.
Condition (ii) is easy to check by using the
relation noted above, $\KR^{-n}(R\sqcup R)=\K^{-n}(R)$,
and the fact that $n$ is even ($p$ odd) for IIB
and $n$ is odd ($p$ even) for IIA.

To test (i), let $W$ be an orientifold $p$-plane.
For the moment, we assume that the normal bundle of $W$ in $X$ is
trivial so that a neighborhood of $W$ in $X$ looks like a product
space $D^{0,n}\times W$ on which $\tau$ acts on the sign-flip of
the disc factor. From our proposal, the group corresponding to
D-branes localized on $W$ can be identified as
$\KR^{-n}(D^{0,n}\times W,\partial D^{0,n}\times W)$
which is the $\KR^{-n}$-group for
bundles over $D^{0,n}\times W$
that vanishes on $\partial D^{0,n}\times W$.
Namely, the KR-group for bundles over
$D^{n,0}\times D^{0,n}\times W$ that vanishes on
$\partial D^{n,0}\times D^{0,n}\times W$ and
$D^{n,0}\times\partial D^{0,n}\times W$, i.e. on
$\partial(D^{n,0}\times D^{0,n})\times W$.
Since $D^{n,0}\times D^{0,n}\cong D^{n,n}$ this is identified with
$\KR(D^{n,n}\times W,\partial D^{n,n}\times W)=
\KR^{-n,-n}(W)$. By the Bott periodicity (\ref{Bott1}),
this is equal to $\KR(W)$ which is indeed $\KO(W)$
since the involution acts trivially on $W$.
Even when the normal bundle is non-trivial, this is true at least
locally. Glueing such local realtions is a generalization of the proof
of Thom isomorphism theorem, and we expect that this is also true
globally.

The proposed group also has another desirable property
that could have been added to the conditions (i) and (ii).
Before showing it, we comment on a related fact
about the condition (i) itself.
When D$(p-4)$-branes are wrapped on
a submanifold $W^{\prime}$ of an O$p$-plane $W$ of $SO$-type,
they support symplectic bundle over $W^{\prime}$.
Thus, D-branes localized in $W^{\prime}$ must be classified by
$\KSp(W^{\prime})$.
Since a neighborhood of $W^{\prime}$ in $W$ locally looks like
$D^4\times W^{\prime}$, the condition (i) shows that
the relevant group is
$\KO(D^4\times W^{\prime},\partial D^4\times W^{\prime})$
which can be identified with
$\KO^{-4}(W^{\prime})$.
This is indeed $\KSp(W^{\prime})$ by Bott periodicity
(\ref{Bott3}).
Now we go back to our proposal.
If $W^{\prime\prime}$ is a $\Z_2$-invariant
submanifold of $X$ of dimension $p+5$ including all the
O$p$-planes (of $SO$-type),
as we have seen in section
3, D$(p+4)$-branes wrapped on $W^{\prime\prime}$
support symplectic orientibundle over $W^{\prime\prime}$.
Thus, D-branes localized in $W^{\prime\prime}$
must be classified by $\KPR(W^{\prime\prime})$.
In fact, a neighborhood of $W^{\prime\prime}$
in $X$ looks locally like $D^{0,5-p}\times W^{\prime\prime}$
and the relevant K-group is identified as
$\KR^{-(9-p)}(D^{0,5-p}\times W^{\prime\prime},
\partial D^{0,5-p}\times W^{\prime\prime})$.
As in the test of (i),
this is equal to $\KR^{-(9-p),-(5-p)}(W^{\prime\prime})$
which is indeed
$\KR^{-4}(W^{\prime\prime})
=\KPR(W^{\prime\prime})$ by Bott periodicity (\ref{Bott1}),
(\ref{Bott3}).

We have shown that our proposal does satisfy all three
requirements ((i), (ii) and another) but we have not shown that
it is the only possibility.
In the next subsection, we will show that
it is the one
that naturally appears
when applying T-duality to Type I string theory
(for $SO$-type orientifold).
\footnote{Chronologically,
that is the way I found the proposal.}

Note that
the proposal does not cover the most general cases.
For example, there are orientifolds including $SO$-type and $Sp$-type
O-planes at the same time.
There could also be cases where there are O-planes of mixed dimensions.
In this paper, we do not attempt to
specify the relevant K-theory group for such exotic (but sometimes
interesting) models.

\subsection{The T-duality Isomorphism}

Now we identify the isomorphism of K-groups
that realizes T-duality on $n$-torus.

We start with the basic example of $n=1$.
Consider a D-brane configuration in Type IIB string theory
on $T^1\times M$ where $M$ is a nine-dimensional manifold.
We represent it as a bound state of D9 and anti-D9-branes
which support Chan-Paton bundles $E^{\pm}$ with
connections $A^{\pm}_M$ and the tachyon field $T:E^+\to E^-$.
We probe it with a D1-brane wrapped on $T^1$.
Then, Kaluza-Klein reduction of the probe theory
is identified with the quantum mechanics of
a D$0$-brane probing the T-dualized system on $\hT^1\times M$.
We coordinatize $T^1$ and $\hT^1$ by $x$ and $a$ respectively.
Mass matrix of the fermions in the quantum mechanics originating
from 1-9 and 1-$\bar 9$ strings are given by the operator
\beq
{\cal D}=\left(\begin{array}{cc}
D_+\!&\!\!-T^{\dag}\\
T&\!-D_-
\end{array}\right).
\label{tachy}
\eeq
Here $D_{\pm}=\partial_x+A^{\pm}_x-ia$ is the
Dirac operator $D$ on $T^1$ coupled to the family of connections
$A^{\pm}_x-ia$ carried by
$E^{\pm}\otimes {\cal P}$ where ${\cal P}$ is the
Poincar\'e bundle over $T^1\times\hT^1$ with curvature
$-i\dd a\wedge\dd x$.
This is interpreted as the tachyon field
of Type IIA ninebranes of the T-dualized system.

For instance, consider (unphysical) Type IIB string theory
on $T^1\times \R^9$ with a single D9-brane with Wilson line
$ia^0\dd x$ where $x$ is the coordinate on $T^1$ with periodicity
$x\equiv x+1$. Then the Dirac operator on $T^1$
for the fermion of the probe D1-brane
is ${\cal D}(a)=\partial_x+ia^0-ia$. Kaluza-Klein modes on $T^1$
consist of functions $\e^{i \hat{n} x}$, $\hat{n}\in 2\pi \Z$,
 with ${\cal D}(a)$ eigenvalues $-i(a-(a^0+\hat{n}))$.
One interpretation of these modes is as the
0-8 strings for a D0-brane probing a D8-brane at the point $a=a^0$
of the dual torus $\hT^1$ (or equivalently
infinite array of D8-branes at $a=a^0+\hat{n}$ in the covering space).
Alternatively, these can also be interpreted as 0-9 strings
for a D0-brane probing Type IIA D9-branes
where ${\cal D}(a)=-i(a-(a^0+\hat{n}))$
are interpreted as
the tachyon fields of D9-branes as appeared in
\cite{Horava} which represent the infinite array of D8-branes.
One can also start with (physical) Type IIB string theory
with D$p$-brane ($p$ is odd say $9-2m$) wrapped on $T^1$
with Wilson line $ia^0\dd x$.
The D$p$-brane can
be represented as a D9-anti-D9 bound state with tachyon field
$T=-i{\bf x}\cdot{\bf\Gamma}$ where ${\bf x}=(x^1,\ldots,x^{2m})$
are the coordinates transverse to the D$p$-brane
and ${\bf\Gamma}=(\Gamma^1,\ldots,\Gamma^{2m})$ 
are the Gamma matrices mapping positive chirality spinors
to negative chirality spinors in $2m$-dimensions.
Then, 1-9 and 1-$\bar 9$ fermions reduce to fermions with
mass matrix
\beq
{\cal D}({\bf x},a)=\left(\begin{array}{cc}
-i(a-(a^0+\hat{n}))&-i{\bf x}\cdot{\bf\Gamma}^{\dag}\\
-i{\bf x}\cdot{\bf \Gamma}&i(a-(a^0+\hat{n}))
\end{array}\right)
\eeq
This can be expressed as
${\cal D}(\vec{x})=-i(\vec{x}-\vec{x}_{\hat{n}})\cdot\vec{\Gamma}$
where $\vec{\Gamma}$ are the Gamma matrices for the
$(2m+1)$-dimensions of $\vec{x}=({\bf x},a)$
and $\vec{x}_{\hat{n}}=({\bf 0},a^0+\hat{n})$.
This can be interpreted as the
tachyon fields of Type IIA D9-branes representing
infinite array of D$(8-2m)$-branes
 located at ${\bf x}={\bf 0}, a=a^0+\hat{n}$.
Thus we have recovered the fact that
a D$p$-brane wrapped on a circle is T-dual to a D$(p-1)$-brane
in the dual circle where the Wilson line on the D$p$-brane
corresponds to the position of the D$(p-1)$-brane \cite{DLP}.

As in $T^4$ case,
the interpretation of the operator ${\cal D}$
as the tachyon
field of ninebranes in the T-dualized system
would lead to an index-theoretic map of K-theory groups
that realizes transformation of D-brane charges under T-duality.
Note that the operator ${\cal D}$
(as well as the Dirac operator $D$ on odd-dimensional space)
is skew-adjoint and therefore its family
index always vanishes as an element of the K-group
of the parameter space.
However, Atiyah and Singer in \cite{ASAd}
\footnote{We thank I.M. Singer for explanation of essential points
which are relevant in the present discussion.}
studied index theory of such families and found that the
index naturally takes values in $\K^{-1}$-group of the parameter
space. This is actually what we wanted because T-duality
maps Type IIB on $T^1\times M$
to Type IIA on $\hT^1\times M$ whose D-brane charges are classified
by $\K^{-1}(\hT^1\times M)$
where $\hT^1\times M$ appears in the present set-up
as the parameter space.
The construction in \cite{ASAd} is as follows. If ${\cal D}(y)$ is
the family of skew-adjoint operators parametrized by $y\in Y$,
we define a family over
$[-{\pi\over 2},{\pi\over 2}]\times Y$ by
\beq
\widetilde{\cal D}(t,y)=-\sin t+{\cal D}(y)\cos t.
\eeq
This is no longer skew-adjoint and
therefore can have distinct kernel and cokernel.
Since
$\widetilde{\cal D}(-{\pi\over 2},y)
=-\widetilde{\cal D}({\pi\over 2},x)\equiv 1$,
kernel and cokernel are isomorphic at $t=\mp{\pi\over 2}$
and therefore the
index of $\widetilde{\cal D}$ belongs to
$\K([-{\pi\over 2},{\pi\over 2}]\times Y,
\partial [-{\pi\over 2},{\pi\over 2}]\times Y)=K^{-1}(Y)$.

The element of $\K^{-1}(\hT^1\times M)$ obtained in this way
in our set-up can be identified as the D-brane charge of
the T-dualized system.
We show this in the example of D$(9-2m)$-brane considered above
where we put $a^0=0$ for simplicity and we compactify
the transverse space $\R^{2m}$ to $S^{2m}$, requiring triviality
at infinity $\infty\in S^{2m}$ (we thank the authors of \cite{BGH}
for pointing out omission of this requirement
in the previous version of the paper).
The operator
$\widetilde{\cal D}(t,\vec{x})=-\sin t+{\cal D}(\vec{x})\cos t$
has zero for constant
modes at $(t,\vec{x})=(0,\vec{0})$,
and there are no zero modes everywhere else.
In a neightborhood of this point, $\widetilde{\cal D}(t,\vec{x})$
on the constant modes behaves as
\beq
\widetilde{\cal D}(t,\vec{x})\,=\,
-\sin t-i\vec{x}\cdot\vec{\Gamma}\cos t\,
\approx\, -(t+i\vec{x}\cdot\vec{\Gamma}).
\eeq
This has winding number one on the $(2m+1)$-sphere surrounding
$(t,\vec{x})=(0,\vec{0})$, and therefore
the index of $\widetilde{\cal D}$
has $\ch_0=\cdots=\ch_{2m}=0$, $\ch_{2m+2}=1$ as an element of
$\K(S^1\times {\hT^1\times S^{2m}\over \hT^1\times \infty})$.
This is the generator of
$\K^{-1}(\hT^1\times S^{2m},\hT^1\times \infty)\cong\Z$
which represents the
charge of the D$(8-2m)$-brane of
the T-dualized system.\footnote{In \cite{Horava},
the element of $\K^{-1}(X)$
corresponding to the (anti-hermiaitin) tachyon field
${\cal T}:E\to E$
of IIA ninebrane is identified as
$(E,\e^{\cal T})$ in the definition
of $\K^{-1}(X)$ in \cite{Karoubi}.
It is $(E_{\e^{\cal T}},E_{\rm id})$
in the standard definition of $\K^{-1}(X)$ as the subgroup of
$\K(S^1\times X)$, where $E_{\varphi}$ for an auatomorphism
$\varphi:E\to E$ is a bundle over $S^1\times X$
defined as $[0,1]\times E$ modulo the identification
$(0,v)\equiv (1,\varphi v)$. In our set-up,
after a suitable regularization of ${\cal T}={\cal D}$,
$\e^{\cal T}$ has winding number one
on a disc $D^{2m+1}$ containing $\vec{x}=\vec{0}$
(i.e. it is constant on $\partial D^{2m+1}$ and
the induced map on $S^{2m+1}=D^{2m+1}/\partial D^{2m+1}$ has winding
number one).
Therefore $E_{\e^{\cal T}}$ and $E_{\rm id}$ differ
in $\ch_{2m+2}$ by one, determining the same element
as the index of $\widetilde{\cal D}$.}

In the framework where the space-time
is compactified by attaching infinity,\footnote{Triviality at infinity
is imposed as usual but, as in \cite{WittenK},
is not explicitly indicated in the general formula
since what ``infinity'' means
depends on (and is clear in) the particular situation one considers.}
the change of the
off-diagonal part of (\ref{tachy})
is a small perturbation and does not affect the index.
Therefore, the index depends only on the K-theory class
of $(E^+\otimes {\cal P},E^-\otimes {\cal P})$
where $E^-\otimes {\cal P}$ gives a negative contribution
because the Dirac operator $D_-$ enters in ${\cal D}$
with minus sign.
In particular, it can be denoted as
$\ind (D,(E^+\otimes {\cal P},E^-\otimes{\cal P}))$.
Thus, the map that realizes
the transformation of D-brane charges under T-duality
can be identified as
\beq
\K(T^1\times M)\,
\stackrel{\otimes{\cal P}}{-\!\!\!\longrightarrow}\,
\K(T^1\times \hT^1\times M)\,
\stackrel{\ind D}{-\!\!\!\longrightarrow}\,
\K^{-1}(\hT^1\times M).
\label{TonT1}
\eeq
In \cite{ASAd} orientibundles were also considered
and it was shown
 that the index of a family of skew-adjoint operators
 on a real Hilbert space
 takes values in $\KR^{-1}$ of the parameter
 space via the same construction.
Thus, we also have
\beq
\KR(T^1\times M)\,
\stackrel{\otimes{\cal P}}{-\!\!\!\longrightarrow}\,
\KR(T^1\times \hT^1\times M)\,
\stackrel{\ind D}{-\!\!\!\longrightarrow}\,
\KR^{-1}(\hT^1\times M),
\label{TonT1R}
\eeq
where the involution acts trivially on $T^1$ but
as the inversion on $\hT^1$.

What if we started from an element of $\K^{-1}(T^1\times M)$?
Such an element can be considered as an element of
$\K(S^1\times T^1\times M)$ that vanishes on $z_0\times T^1\times M$.
Tensoring with ${\cal P}$ and taking the index
do not affect the triviality
at $z_0\in S^1$. Thus, its image
under the map (\ref{TonT1}), which a priori 
belongs to
$\K^{-1}(S^1\times\hT^1\times M)$, actually belongs to the subgroup
$\K^{-2}(\hT^1\times M)$. Such a consideration leads to the following
generalizations of (\ref{TonT1}) and (\ref{TonT1R})
\beqa
&&\K^{-i}(T^1\times M)\,
\stackrel{\otimes{\cal P}}{-\!\!\!\longrightarrow}\,
\K^{-i}(T^1\times \hT^1\times M)\,
\stackrel{\ind D}{-\!\!\!\longrightarrow}\,
\K^{-i-1}(\hT^1\times M),\\
&&\KR^{-i}(T^1\times M)\,
\stackrel{\otimes{\cal P}}{-\!\!\!\longrightarrow}\,
\KR^{-i}(T^1\times \hT^1\times M)\,
\stackrel{\ind D}{-\!\!\!\longrightarrow}\,
\KR^{-i-1}(\hT^1\times M).
\eeqa
By composition, we obtain maps
\beqa
&&
\K^{-i}(T^n\times M)\longrightarrow \K^{-i-n}(\hT^n\times M),
\label{TonTn}\\
&&\KR^{-i}(T^n\times M)\longrightarrow \KR^{-i-n}(\hT^n\times M),
\label{TonTnR}
\eeqa
representing T-duality on $n$-torus
where now $M$ is a $(10-n)$-dimensional manifold.
In the latter case, the involution acts trivially on $T^n$
and as the inversion on $\hT^n$ while
$M$ is assumed to have
a suitable involution
so that $i$ can be chosen appropriately.

As a consistency check,
let us consider
Type I string on $T^n\times M$ which T-dualizes to
Type II orientifold on $\hT^n/\Z_2\times M$ of $SO$-type.
Type I D-brane charge takes values in
$\KO(T^n\times M)=\KR(T^n\times M)$. Therefore,
by (\ref{TonTnR}),
we see that D-brane charges of Type II orientifold
on $\hT^n/\Z_2\times M$ must be classified by
$\KR^{-n}(\hT^n\times M)$, thus reproducing our proposal
(for $SO$-type orientifold).
Also, when involution acts on $M$
with only $\R^s/\Z_2$-type fixed points,
according to our proposal,
D-brane charges of orientifold on $T^n\times M/\Z_2$
are classified by $\KR^{-i}(T^n\times M)$ with
$i=s$ or $s+4$ for $SO$- or $Sp$-type respectively.
Then (\ref{TonTnR}) shows that D-branes in orientifold
on $(\hT^n\times M)/\Z_2$ is classified by
$\KR^{-j}(\hT^n\times M)$ with $j=s+n$ or $s+n+4$, which is consistent
with our proposal since there are only $\R^{n+s}/\Z_2$ fixed points.

Since T-duality squares to the identity, the maps we have obtained
must be isomorphisms.
Here we do not attempt to compute these isomorphisms.
Instead, we show that the groups are indeed isomorphic.

\subsection*{\sl K-theory Groups of $T^n\times M$}

A subspace $B$ of a topological space $A$ is said to be
a {\it retract}
of $A$ if there is a continuous map $f:A\to B$ that restricts on $B$
to the identity. In such a case, there is a relation
$\K(A)=\K(A,B)\oplus \K(B)$ which holds also for
$\K^{-i}$ and $\KR^{-p,-q}$.
Now let us apply this to $z_0\times M\subset T^1\times M$
which is obviously a retract.
Noting that
$K^{-i}(T^1\times M, z_0\times M)=K^{-i-1}(M)$,
we have
\beq
\K^{-i}(T^1\times M)=\K^{-i-1}(M)\oplus \K^{-i}(M).
\eeq
To consider the case with invoultion,
as before we let $\Z_2$ act on $T^n$ trivially and
on $\hT^n$ as the inversion.
Then, $z_0\times M\subset T^1\times M$ and
$\hat{z}_0\times M\subset \hT^1\times M$ are both retracts
where $\hat{z}_0$ is a $\Z_2$ fixed point.
Noting that
$\KR^{-p,-q}(T^1\times M,z_0\times M)=\KR^{-p-1,-q}(M)$
and $\KR^{-p,-q}(\hT^1\times M,\hat{z}_0\times M)
=\KR^{-p,-q-1}(M)$, we have
\beqa
&&\KR^{-p,-q}(T^1\times M)
=\KR^{-p-1,-q}(M)\oplus \KR^{-p,-q}(M),\\
&&\KR^{-p,-q}(\hT^1\times M)
=\KR^{-p,-q-1}(M)\oplus \KR^{-p,-q}(M).
\eeqa
Using these relations repeatedly,
we obtain the binomial identities
\beqa
&&\K^{-i}(T^n\times M)=\bigoplus_{k=0}^n\,
{\vbox to 10pt{}\K^{-i-k}(M)}^{\oplus{n\choose k}},
\label{bino1}\\
&&\KR^{-p,-q}(T^n\times M)=\bigoplus_{k=0}^n\,
{\vbox to 10pt{}\KR^{-p-k,-q}(M)}^{\oplus{n\choose k}},
\label{bino2}\\
&&\KR^{-p,-q}(\hT^n\times M)=\bigoplus_{k=0}^n\,
{\vbox to 10pt{}\KR^{-p,-q-k}(M)}^{\oplus{n\choose k}},
\label{bino3}
\eeqa
where $M$ is any space (with an involution in the latter two).
Using Bott periodicity $\K^{-j-2}=\K^{-j}$,
$\KR^{-p-1,-q-1}=\KR^{-p,-q}$, we see from these that
\beqa
&&\K^{-i}(T^n\times M)\,\cong\,\K^{-i-n}(\hT^n\times M),\\
&&\KR^{-i}(T^n\times M)\,\cong\,\KR^{-i-n}(\hT^n\times M).
\eeqa
Thus, the groups in the left hand side and the right hand side 
of the T-duality maps (\ref{TonTn}) and (\ref{TonTnR}) are
indeed isomorphic.

Applying the binomial identities (\ref{bino1})-(\ref{bino3})
to the case where $M$ is a point,
let us compute some K-groups of torus.
We first note that $\K({\rm pt})=\Z$ and $\K^{-1}({\rm pt})=0$.
Then , we immediately obtain from (\ref{bino1})
\beq
\K(T^n)\cong{\vbox to 10pt{}\Z}^{\oplus 2^{n-1}},~~~
\K^{-1}(T^n)\cong{\vbox to 10pt{}\Z}^{\oplus 2^{n-1}}.
\label{Ktor}
\eeq
These are mapped isomorphically by the
Chern character maps
$\ch:\K(X)\to H^{\rm even}(X,\Q)$,
$\K^{-1}(X)\to H^{\rm odd}(X,\Q)$
to the even and odd dimensional
integral cohomology groups of the torus respectively
(or more precisely their images in the rational cohomology).
Note that $2^{n-1}$ is the dimension of the (positive or negative)
spinor representation of $SO(2n)$.
In fact, it is known that the T-duality group $O(n,n;\Z)$
acts on the RR potentials in the spinor representation,
where IIA potentials belong to, say, positive spinor
representation
while IIB potentials belong to negative spinor
representation which are interchanged under odd number of
T-duality on circles \cite{OP}.
This fact may be explained by showing that
our T-duality maps for D-brane charges (together with
the mapping class group of $T^n$ and
some operation corresponding B-field shift)
generate spinor representation of $O(n,n;\Z)$.

To compute $\KR^{-i}(T^n)$, we need to know
$\KR^{-i}({\rm pt})=\KO^{-i}({\rm pt})$. Non-zero groups of them
are
$\KO({\rm pt})=\Z$, $\KO^{-1}({\rm pt})=\Z_2$,
$\KO^{-2}({\rm pt})=\Z_2$ and $\KO^{-4}({\rm pt})=\Z$.
Inserting these into the identities (\ref{bino2}) and (\ref{bino3})
we obtain for example
\beqa
&&\KO(T^4)=\Z\oplus\Z_2^{\oplus 4}\oplus\Z_2^{\oplus 6}\oplus
\Z=\KPR(\hT^4),\\
&&\KSp(T^4)=\Z\oplus\Z=\KR(\hT^4).
\eeqa
It is intereting to investigate what the full T-duality group is
and in which representation the D-brane charges (and RR fields)
belong to.

\appendix{The Inversion Theorem}

In this appendix we show that square Nahm transform is the identity
operation
for orthogonal/symplectic (orienti)bundles over $T^4$ ($T^4/\Z_2$)
or for unitary orbibundles over orbifold $T^4/\Z_2$.
Namely, we prove that there is an isomorphism
\beq
(E,A,\Sigma)\stackrel{\cong}{\longrightarrow}
(E,A,\Sigma)\hat{\vbox to 7pt{}}\hspace{0.14cm}\check{\vbox to 8pt{}}
\label{isom}
\eeq
where $\Sigma$ is either one of
\vspace{-0.3cm}
\begin{itemize}
\setlength{\itemsep}{-0.2pt}
\item[(i)] the orthogonal structure
defined by anti-linear maps $I_x:E_x\to E_x$
\item[(ii)] the symplectic structure
defined by anti-linear maps $J_x:E_x\to E_x$
\item[(iii)] the $\Z_2$-orbifold action defined by
linear maps $\varphi_x:E_x\to E_{-x}$
\end{itemize}

\vspace{-0.3cm}
\noindent
on the unitary bundle $E$ over $T^4$, and ``~$\hat{}$~'' 
and ``~$\check{}$~'' are the Nahm
transforms
which we have constructed in sections 3
(~$\check{}$~ is equal to ~$\hat{}$~ for the case (iii)).
It is already known that there is an isomorphism
$(E,A)\cong 
(E,A)\hat{\vbox to 6pt{}}\hspace{0.14cm}\check{\vbox to 7pt{}}\,$
\cite{Mu,BvB} (see also\cite{DK}) if we ignore the respective structure
$\Sigma$.
Therefore, what we need to do is to show that this
isomorphism sends the structure $\Sigma$ to that of the
square Nahm transform.
We present the proof for the case (i) in some detail, and only
indicate the essential points for the cases (ii) and (iii).

We first recall the definition of the isomorphism $u:E\to
\hat{E}\check{\vbox to 9pt{}}$ \cite{BvB}.
Let us denote by $G_a$ the inverse of the Laplace operator
$D^{(a)\dag}D^{(a)}$ associated with the covariant derivative
$D^{(a)}_{\mu}=\partial_{\mu}+A_{\mu}-ia_{\mu}$.
Then for $v\in E_x$,
$u(v)\in \Bigl(
\hat{E}\check{\vbox to 9pt{}}\,
\Bigr)_x$
is given as the section of the bundle $\hat{E}\otimes \hat{S}^+$
defined by
\beq
u(v): a\longmapsto \sqrt{-1}[\psi^{(a)}_i]\otimes C((G_a\psi_i^{(a)})(x),v)
\label{udef}
\eeq
where 
$\{\psi_i^{(a)}\}$ is an orthonormal basis
of the kernel of the Dirac operator $\Dsl^{(a)}$ (representing
an orthonormal basis $\{[\psi_i^{(a)}]\}$
of $\hat{E}_a$),
$C$ is the charge conjugation matrix and $(\,,\,)$ is the hermitain
metric of $E$ which is anti-linear in the left-entry and linear in the
right-entry.
(The phase $\sqrt{-1}$ in front is simply for later convenience.)
To be precise, (\ref{udef}) defines a section of
$\hat{E}\otimes {\cal L}_x\otimes \hat{S}^+$ where ${\cal L}_x$
is a flat bundle defined as
the quotient of $(\R^4)^*\times \C$ with the trivial connection
by the $2\pi \Lambda^*$ action
$(a,c)\mapsto (a+\hat{n},\e^{-i\hat{n}x}c)$.
(This is because $\psi^{(a)}(x)\in \Ker\Dsl^{(a)}$ and 
$\psi^{(a+\hat{n})}(x)=\e^{i\hat{n}x}\psi^{(a)}(x)\in
\Ker\Dsl^{(a+\hat{n})}$ are identified as an element of
$\hat{E}_a$.)
However, ${\cal L}_x$ is topologically trivial and is isomorphic
as a flat bundle to
the trivial bundle $\hT^4\times \C$ with the connection
$D=d-ix^{\mu}\dd a_{\mu}$.
By this identification, we consider $u(v)$ as a section of
$\hat{E}\otimes \hat{S}^+$.
One can show, as in \cite{BvB}, that
$u(v)$ is annihilated by
the Dirac operator associated with the covariant derivative
$\partial/\partial a_{\mu}+\hat{A}^{\mu}-ix^{\mu}$,
thus showing that $u(v)$ belongs to
$\Bigl(\hat{E}\check{\vbox to 9pt{}}\,\Bigr)_x$.
Moreover, it was shown in \cite{BvB} that $u$ sends the hermitian
metric of $E$ to that of $\hat{E}\check{\vbox to 9pt{}}$
and the connection $A$ to $\hat{A}\check{\vbox to 9pt{}}$.

Now let us show that $u$ sends the orthogonal structure $I$ of
$(E,A)$
to that $\check{I}$ of
$(E,A)\hat{\vbox to 6pt{}}\hspace{0.14cm}\check{\vbox to 7pt{}}$.
Namely we show that
\beq
\check{I}_xu(v)=u(I_xv),
\label{invth}
\eeq
for each $v\in E_x$. By definition, we have
$\check{I}u(v)\Bigr|_a=J^c_{-a}(u(v)|_{-a})$, and therefore
\beqa
\check{I}u(v)\Bigr|_a&=&
-\sqrt{-1}J_{-a}([\psi_i^{(-a)}])\otimes
\gamma^{1234} [C((G_{-a}\psi_i^{(-a)})(x),v)]^c
\nonumber\\
&=&
\sqrt{-1}
[I^c\psi_i^{(-a)}]\otimes C\overline{C}(v,(G_{-a}\psi_i^{(-a)})(x))
\nonumber\\
&=&
\sqrt{-1}
[I^c\psi_i^{(-a)}]\otimes
C\overline{C}(I_x(G_{-a}\psi_i^{(-a)})(x),I_xv)\nonumber\\
&=&
\sqrt{-1}
[I^c\psi_i^{(-a)}]\otimes
C(C(IG_{-a}\psi_i^{(-a)})(x),I_xv),\label{A7}
\eeqa
where $I$ is considered as an operator
acting on the section $s(x)$ of $E$ as $(Is)(x):=I_xs(x)$.
Note that $IG_{-a}=G_aI$ which follows from
$ID^{(a)}=D^{(-a)}I$.
Then, we see that
$$
C(IG_{-a}\psi^{(-a)})(x)=
CG_aI\psi^{(-a)}(x)=G_aI^c\psi^{(-a)}(x).
$$
Inserting this to (\ref{A7}), we have shown (\ref{invth}):
\beqa
\check{I}u(v)\Bigr|_a&=&
\sqrt{-1}
[I^c\psi_i^{(-a)}]\otimes
C((G_aI^c\psi_i^{(-a)})(x),I_xv)\nonumber\\
&=&u(I_xv)|_a,
\eeqa
thus proving (\ref{isom}).

The proof for the cases of (ii) and (iii) are similar.
The essential points in these cases are
$JG_{-a}=G_aJ$ for (ii) and $\varphi G_{-a}=G_a\varphi$ for (iii),
where $J$ and $\varphi$ are considered as an operator acting on the
section $s$ of $E$ as $(Js)(x)=J_xs(x)$ and
$(\varphi s)(x)=\varphi_{-x}s(-x)$ respectively.

\section*{Acknowledgements}

I would like to thank B. Acharya, J. Blum,
J. de Boer, S. Kachru, A. Karch, J. Maldacena,
N. Obers, H. Ooguri, A. Strominger, C. Vafa and P. Yi
for useful discussions
and M. Hopkins and I.M. Singer for help and instructions in
K-theory and index theory.
I wish to thank Aspen Center for Physics,
Physics and Mathematics Departments of
Harvard University, and Korea Institute for Advanced Study,
where various stages of this work
were carried out, for their kind hospitality.
I express my thanks and apologies to many people
who kindly
pointed out omission
and incorrect citation in the previous version of the paper.

This research is supported in part by NSF grant PHY-95-14797
and DOE grant DE-AC03-76SF00098.

\end{document}